\documentclass{article}
\usepackage[english]{babel}
\usepackage[a4paper, top=2.5cm, bottom=2.5cm, left=2.5cm, right=2.5cm]{geometry}
\usepackage{chngcntr}
\usepackage{amsmath}
\usepackage{float}
\usepackage{graphicx}
\usepackage{booktabs}
\usepackage{siunitx}
\usepackage{authblk}
\usepackage{orcidlink}
\usepackage[authoryear,round]{natbib}
\PassOptionsToPackage{colorlinks=true, allcolors=blue}{hyperref}
\usepackage{hyperref}

\title{Reduced NEXI protocol for the quantification of human gray matter microstructure on the Connectome 2.0 scanner}

\author[1]{Quentin Uhl \orcidlink{0000-0002-9160-4605}}
\author[1]{Tommaso Pavan \orcidlink{0000-0002-3436-6882}}
\author[2]{Julianna Gerold \orcidlink{0009-0000-5460-6591}}
\author[2]{Kwok-Shing Chan \orcidlink{0000-0001-8427-169X}}
\author[2]{Yohan Jun \orcidlink{0000-0003-4787-4760}}
\author[2]{Shohei Fujita \orcidlink{0000-0003-4276-6226}}
\author[2]{Aneri Bhatt \orcidlink{0009-0003-3502-9399}}
\author[2]{Yixin Ma \orcidlink{0000-0002-2080-9013}}
\author[1]{Qiaochu Wang \orcidlink{0009-0007-0234-6864}}
\author[2]{Hong-Hsi Lee \orcidlink{0000-0002-3663-6559}}
\author[2]{Susie Y. Huang \orcidlink{0000-0003-2950-7254}}
\author[2,*]{Berkin Bilgic \orcidlink{0000-0002-9080-7865}}
\author[1,*]{Ileana Jelescu \orcidlink{0000-0002-3664-0195}}

\affil[1]{Department of Radiology, Lausanne University Hospital (CHUV) and University of Lausanne, Lausanne, Switzerland}
\affil[2]{Department of Radiology, Massachusetts General Hospital, Athinoula A. Martinos Center for Biomedical Imaging, Boston, MA, USA}
\affil[*]{Joint last authorship}

\begin{document}

\maketitle

% ========= Corresponding author =========
\begin{center}
\vspace{-1em}
Corresponding author: Quentin Uhl (\href{mailto:quentin.uhl@gmail.com}{quentin.uhl@gmail.com})
\end{center}
\vspace{1.5em}
% =======================================================

% ========= Keywords =========
\vspace{-1em}
\vspace{1.5em}

% ========= Abstract =========
\vspace{-1em}
\noindent
\textbf{Abstract}: Biophysical diffusion MRI models like Neurite Exchange Imaging (NEXI) are essential for probing gray matter microstructure, estimating compartment diffusivities ($D_i, D_e$), neurite fraction ($f$), and exchange time ($t_{ex}$). However, NEXI multi-shell, multi-diffusion-time requirements lead to prohibitively long acquisitions. Leveraging the Connectome 2.0 ultra-high gradient scanner, we developed a time-efficient protocol using an Explainable AI (XAI) framework. Combining XGBoost, SHAP, and Recursive Feature Elimination trained on synthetic signals, XAI identified an optimal 8-feature subset, halving scan time from 27 to 14 minutes. Validated in vivo in seven healthy participants, the XAI protocol was benchmarked against the full 15-feature acquisition, a Cramér-Rao Lower Bound (CRLB) theoretical optimum, and two expert-designed heuristics ("Mid-Range" and "Corner"). The XAI protocol robustly reproduced parameter estimates and maintained full-protocol test-retest reproducibility. Remarkably, the data-driven XAI selection converged to the theoretical CRLB optimum. This validates XAI mathematical optimality while highlighting its primary advantage: achieving gold-standard optimization without requiring complex analytical Jacobians, making it effortlessly adaptable to numerically defined models or complex noise distributions where CRLB becomes intractable. Furthermore, XAI demonstrated superior in vivo robustness compared to heuristics: "Mid-Range" sampling yielded biased $t_{ex}$ estimates due to insufficient temporal diversity, while "Corner" sampling yielded unstable $D_i$ estimates (5-fold higher coefficient of variation) due to noise sensitivity. Ultimately, this robust 14-minute NEXI protocol accelerates exchange-sensitive microstructural mapping, establishing a generalizable, model-agnostic optimization framework adaptable to future ultra-high gradient systems (e.g., Siemens Cima.X or GE MAGNUS) and existing clinical scanners.

\vspace{1.5em}

% ========= Introduction =========
\section{Introduction}

Mapping cortical microstructure is a growing priority in neuroscience and clinical imaging, particularly for detecting early changes in neurodegenerative diseases like Alzheimer's disease, Parkinson's disease and multiple sclerosis, or in neuropsychiatric disorders such as schizophrenia \citep{dongDiffusionMRIBiomarkers2020, atkinson-clementDiffusionTensorImaging2017, spotornoMeasuresCorticalMicrostructure2023}. These pathologies are often marked by subtle alterations in gray matter cytoarchitecture, such as neurite beading, reduced spine or dendritic density, synaptic loss and changes in membrane permeability, that precede detectable volumetric atrophy. Advanced diffusion MRI (dMRI) techniques now enable indirect but noninvasive quantification of these microstructural processes by modeling the mean squared displacement of water molecules in tissue \citep{basserEstimationEffectiveSelfdiffusion1994, lebihanImagingDiffusionMicrocirculation1991}.

Biophysical dMRI models have extended the utility of diffusion imaging beyond empirical metrics to estimate biologically relevant parameters such as neurite density, intra- and extra-cellular diffusivities, exchange rates, and soma size \citep{alexanderImagingBrainMicrostructure2019, jelescuChallengesBiophysicalModeling2020, novikovQuantifyingBrainMicrostructure2019, staniszAnalyticalModelRestricted1997}. 

However, the assumptions of the early models, largely developed for the aligned axonal structures of white matter, are challenged by the complex cytoarchitecture of cortical gray matter. Several model extensions have emerged to account for these effects, including SANDI \citep{palomboSANDICompartmentbasedModel2020, palomboAbundanceCellBodies2018}, which incorporates a spherical soma compartment; and NEXI \citep{jelescuNeuriteExchangeImaging2022}, which models exchange between intra- and extra-neurite spaces. A parallel formulation of the same model, SMEX \citep{olesenDiffusionTimeDependence2022}, accounts also for finite gradient durations by solving the exchange equations numerically, which becomes essential when the narrow pulse approximation (NPA) breaks down, such as on standard clinical scanners with weaker gradients.

NEXI is particularly appealing for gray matter applications: by explicitly modeling exchange, it estimates an apparent exchange time $t_{ex}$, alongside intra-neurite diffusivity $D_i$, extra-neurite diffusivity $D_e$, and the intra-neurite volume fraction $f$. Unlike SANDI, NEXI does not explicitly model the soma contributions, instead absorbing them into the extra-neurite compartment, a choice supported by evidence that exchange dominates signal evolution at standard diffusion times ($>$20 ms) \citep{leeVivoObservationBiophysical2020, olesenDiffusionTimeDependence2022, jelescuNeuriteExchangeImaging2022}. Models such as SANDIX \citep{olesenDiffusionTimeDependence2022}, which incorporate both exchange and spherical restriction, offer a more comprehensive microstructural picture, but this richness comes at the cost of increased model complexity. Estimating a large number of parameters simultaneously can lead to unstable fits and unreliable estimates \citep{olesenDiffusionTimeDependence2022}.

Recent studies on 3 Tesla Connectome 1.0 and clinical-grade scanners have shown that NEXI and SMEX can be implemented in humans, and that the estimated exchange time correlates strongly with myelin-sensitive imaging metrics such as the myelin water fraction (MWF) \citep{uhlQuantifyingHumanGray2024, uhlHumanGrayMatter2025}. These findings establish $t_{ex}$ as a potential biomarker of membrane permeability and myelination, with spatial maps revealing biologically plausible patterns across the cortex. However, the underlying acquisition protocols are lengthy (typically $>$30 minutes) as they require extensive sampling across both gradient amplitude and diffusion time dimensions (i.e. the acquisition of multi-b, multi-t data). The long acquisition time limits the scalability of NEXI in clinical and research settings where scan time, subject motion and compliance are critical constraints \citep{spotornoMeasuresCorticalMicrostructure2023, illan-galaCorticalMicrostructurePrimary2022}.

The introduction of the Connectome 2.0 scanner \citep{huangConnectome20Developing2021, ramos-llordenUltrahighGradientConnectomics2025} offers new opportunities for advanced dMRI protocol design. With 500 mT/m gradients and 600 T/m/s slew rates, it enables high b-values and short diffusion times (achieved with short gradient pulse separations, $\Delta$), essential for probing fast exchange and structural disorder \citep{chanVivoHumanNeurite2025}. Furthermore, the enhanced hardware performance provides a robust benchmark for validating estimations from clinical scanners with lower gradient strengths. On this platform, the standard NEXI model becomes more viable, as gradient pulse durations ($\delta = 5$ ms) approach the idealized NPA, reducing the need for pulse-width corrections \citep{chanVivoHumanNeurite2025}. Yet, even on Connectome 2.0, total acquisition time remains a bottleneck. Reducing the number of ($b$, $\Delta$) combinations while preserving parameter accuracy is therefore essential for clinical translation.

Historically, the design of dMRI acquisition schemes has relied on "top-down" analytical frameworks aimed at maximizing the sensitivity of the signal to specific physical phenomena. Early efforts focused on the optimization of Diffusion Tensor Imaging (DTI) by determining the minimum number of gradient directions and optimal b-values required for robust anisotropy measurements \citep{jonesOptimalStrategiesMeasuring1999}. This matured into a generalized framework for experimental design based on the Fisher Information Matrix (FIM) and the Cramér-Rao Lower Bound (CRLB), which seeks to minimize parameter estimation variance \citep{alexanderGeneralFrameworkExperiment2008, pootOptimalExperimentalDesign2010}. Such model-based approaches have been successfully applied to white matter models \citep{coelhoReproducibilityStandardModel2022} and more recently extended to multidimensional dMRI encoding \citep{bocquillonOptimizationAcquisitionSchemes2025}, where dictionary-based greedy algorithms or CRLB-informed selection are used to navigate the vast parameter space of b-tensor shapes.

However, strictly relying on CRLB-based analytical optimization presents significant practical limitations and is not easily generalizable. Formulating the required analytical Jacobians is mathematically cumbersome or even intractable for advanced models accounting for complex effects, such as inter-compartmental exchange or restricted diffusion in non-idealized geometries. Furthermore, highly non-linear models like NEXI or SMEX can be extremely sensitive to these Jacobian derivations, and the standard theoretical CRLB framework struggles to easily incorporate complex, non-Gaussian noise distributions \citep{karlsenParameterEstimationRiciandistributed1999} or hard physical boundary constraints. These mathematical bottlenecks have motivated the emergence of "bottom-up" data-driven strategies that identify the most informative measurements without prior assumptions about the underlying distribution \citep{grussuFeasibilityDataDrivenModelFree2021}. Recent advances in physics-informed machine learning have demonstrated that neural networks, such as Concrete Autoencoders, can be used to perform differentiable feature selection, effectively pruning comprehensive datasets to find optimal measurement subsets \citep{balinConcreteAutoencodersDifferentiable2019, planchuelo-gomezOptimisationQuantitativeBrain2024}. While these methods excel at reconstructing the full signal from sparse samples, their primary focus is often signal prediction rather than the direct, interpretable quantification of biophysical parameters.

Despite these advances, a gap remains between purely theoretical CRLB optimization, which is locally defined and sensitive to the choice of a ground-truth "anchor" point, and deep-learning-based selection, which often operates as a "black box." Explainable AI (XAI) offers a bridge by providing a statistically rigorous yet interpretable ranking of features based on their contribution to the final estimate. By leveraging global sensitivity analysis via SHAP values, XAI-driven optimization captures the non-linear degeneracies inherent in models like NEXI while maintaining a clear link to the physical acquisition dimensions ($b, \Delta$). This framework allows for a direct comparison with both heuristic expert designs and classical A-optimality, ensuring that the resulting protocol is not only efficient but also physically grounded and clinically reproducible.

To bridge this gap, we implement an XAI-oriented protocol reduction framework built around three key components: gradient-boosted decision trees (XGBoost) \citep{chenXGBoostScalableTree2016}, SHapley Additive exPlanations (SHAP) \citep{lundbergUnifiedApproachInterpreting2017}, and Recursive Feature Elimination (RFE). 

While the signals used for training are synthetic, this approach follows a "bottom-up" data-driven logic where the importance of each acquisition feature is learned from the statistical observation of one million noisy signal instances. By training on such an extensive dataset, the framework captures the intricate non-linear relationships and degeneracies inherent to the NEXI model under realistic noise conditions, rather than relying solely on local analytical derivations. 

However, the complexity and opacity of many modern models, particularly gradient-boosted ensembles, can hinder interpretability, which is essential for clinical trust and scientific insight. XGBoost is especially well-suited to this task due to its high performance on tabular data and robustness to noise, but while individual decision trees are interpretable, large ensembles are complex. We address this using fast post-hoc explanation methods \citep{sagiApproximatingXGBoostInterpretable2021}, where SHAP is the key component. By leveraging cooperative game theory, SHAP assigns consistent importance values to each acquisition condition by accounting for the interaction between features. Each unique combination of b-value and diffusion time is considered a feature, which RFE then progressively prunes based on these SHAP rankings. Together, this XGBoost-SHAP-RFE pipeline, hereafter referred to as XAI, implements an explainable, scalable approach to acquisition design, with successful applications in medical imaging \citep{zhuIntegratedApproachFeature2025, rufinoPerformanceExplainabilityFeature2024}, neuroscience \citep{huangMultilayerStackingMethod2024}, and even non-biomedical domains such as environmental monitoring \citep{wangCouplingInterpretableFeature2025}, cybersecurity \citep{ahmedHybridBaggingBoosting2024}, and user modeling \citep{chernyaevaPredictionCustomerSatisfaction2023}. It is noteworthy that other approaches such as polynomial regression are also interpretable \citet{coelhoAssessmentPrecisionAccuracy2024}; however, XAI handles non-linear interactions and noise distributions without assuming a specific functional form for the error surface.

In this study, we extend this approach to diffusion MRI. We train XGBoost models on synthetic NEXI signals and use SHAP-based RFE to identify the smallest protocol subset that preserves estimation accuracy for $t_{ex}$, $f$, $D_i$, and $D_e$. We benchmark this protocol against a theoretically optimal Weighted CRLB design and two expert-designed heuristic baselines (representing standard 'Mid-Range' and extreme 'Corner' sampling strategies). We validate it in vivo on Connectome 2.0, and assess its fidelity, reproducibility, and adaptability to higher-resolution imaging. This resolution, while not at the absolute limit of what is possible, is considered high for complex diffusion models and represents a significant step toward addressing partial volume effects in cortical gray matter. Our results show that simulation-based data-driven optimization not only reduces scan time to 14 minutes (vs 27 minutes for the full protocol, and 30-40 minutes for previous NEXI protocols \cite{uhlNEXIQuantificationHuman2024, uhlHumanGrayMatter2025}), but also converges with the theoretical optimum while significantly outperforming heuristic designs in preserving biologically relevant contrast and parameter stability.

Ultimately, this work highlights the promise of interpretable machine learning for advancing diffusion model design and making advanced microstructural imaging more accessible in both research and clinical contexts.

% ========= Methods =========
\section{Methods}

\subsection{Participants}

Seven healthy adult volunteers (3 males, mean age 30.3 ± 4.0 years) were scanned under informed consent in accordance with institutional ethical guidelines. Each participant underwent two identical scan sessions on the Connectome 2.0 MRI system, performed one after the other, to evaluate test-retest reproducibility of NEXI parameter estimation.

\subsection{MRI acquisition}

All imaging was performed on a MAGNETOM Connectom.X 3T scanner (Siemens Healthineers) equipped with a 500 mT/m gradient amplitude and 600 T/m/s slew rate \citep{ramos-llordenUltrahighGradientConnectomics2025}, using a 72-channel head coil \citep{mahmutovicHighdensityMRICoil2025}. Diffusion-weighted imaging (DWI) was acquired with a single-shot spin-echo EPI sequence and monopolar gradient pulses. Imaging parameters were: voxel size = 2 mm isotropic, Repetition Time (TR)/Echo Time (TE) = 3300/66 ms, matrix size = 96 × 96, 72 axial slices, partial Fourier = 6/8, in-plane acceleration = 2 and simultaneous multislice (SMS) acceleration = 2.

The full protocol was a comprehensive acquisition scheme designed heuristically to ensure broad sampling of the acquisition parameter space. It consisted of 15 ($b$, $\Delta$) combinations, spanning b-values from 1.0 to 12.5 ms/µm$^2$ and diffusion gradient separations $\Delta$ from 12 to 45 ms, with a fixed gradient duration $\delta = 5$ ms (where the effective diffusion time is defined as $t_d=\Delta-\delta/3$). The restriction to relatively short diffusion times ($\Delta \leq 45$ ms) is driven by the trade-off between exchange sensitivity and signal-to-noise ratio (SNR). Extending $\Delta$ necessitates a longer TE, which drastically reduces SNR due to $T_2$ relaxation and compromises the stability of the fit. A maximum $\Delta$ of $45$ ms was selected as the optimal compromise. Directional sampling ranged from 20 to 64 directions depending on b-value (20 directions at $b=1$, 30 at $b=2.5$, 32 at $b=5$, 34 at $b=7.35-7.50$, 44 at $b=10.0$, and 64 at $b=12.5$ ms/µm$^2$). The diffusion time needs to be fixed for each diffusion MRI protocol. Data acquisition was thus split across five different sequences, one for each $\Delta$. Each scan also included non-diffusion-weighted images ($b=0$), totaling 13 volumes. The number of diffusion directions per shell was scaled heuristically with the b-value (approximately following a square-root dependence, $N_{dir} \propto \sqrt{b}$) to partially compensate for the expected drop in SNR at higher diffusion weightings, thereby harmonizing the SNR across shells. Non-diffusion-weighted images ($b=0$) were acquired at each distinct $\Delta$ value to serve as specific reference signals, allowing for block-wise normalization to mitigate potential signal drifts or inter-sequence motion. Total scan time was approximately 27 minutes (see Table \ref{tab:protocol_details}).

% ---------- Table 1 ----------
\begin{table}[ht]
\centering
\caption{\textbf{Details of the full acquisition protocol.} Summary of b-values, number of diffusion directions per shell ($N_{\text{dir}}$), and associated diffusion times ($\Delta$). The gradient duration is fixed at $\delta = 5$ms for all acquisitions.}
\label{tab:protocol_details}
\begin{tabular}{|c|c|c|}
\hline
\textbf{b-value} & \textbf{Directions} & \textbf{Diffusion time} \\
(ms/\textmu m$^2$) & ($N_{\text{dir}}$) & $\Delta$ (ms) \\
\hline
1.0 & 20 & 12, 27, 45 \\
\hline
2.5 & 30 & 12, 27, 45 \\
\hline
5.0 & 32 & 15, 27, 45 \\
\hline
7.35 / 7.5 & 34 & 20, 27, 45 \\
\hline
10.0 & 44 & 27, 45 \\
\hline
12.5 & 64 & 45 \\
\hline
\multicolumn{3}{|l|}{\textit{Note:} $b=0$ images were acquired for each $\Delta$.} \\
\hline
\end{tabular}
\end{table}

This specific experimental design formulates the optimization problem as a discrete subset selection task (feature elimination) from a master protocol, rather than a continuous optimization of ($b$, $\Delta$) coordinates. While continuous optimization theoretically allows for finer tuning, implementing arbitrary unique $\Delta$ values on clinical scanners typically introduces significant logistical overhead, often requiring separate sequence executions for each timing. By optimizing a subset of a feasible master protocol, we ensure immediate translational applicability.

The reduced protocol retained 8 features selected via a SHAP-based feature elimination strategy (see below), with 250 directions in total and identical $b=0$ sampling. Acquisition time for the reduced protocol was 14 minutes. The 8 selected features spanned a broad range of b-values and diffusion times: (b,$\Delta$) pairs of (1.0, 12), (2.5, 12), (5.0, 15), (7.3, 20), (1.0, 27), (1.0, 45), (2.5, 45) and (12.5, 45) ms/µm$^2$ and ms, respectively. Both protocols were pre-processed and processed independently to allow for a matched comparison.

To enable cortical segmentation, a high-resolution T1-weighted anatomical scan (MPRAGE) was collected during each session with 0.9 mm isotropic resolution (TR/TE = 2500/2.9 ms, TI = 1100 ms, flip angle = 8°). 

To preliminarily assess the protocol's robustness to partial volume effects, an additional higher-resolution acquisition was performed using the reduced 8-feature protocol at 1.6 mm isotropic resolution for one last participant. To compensate for the lower SNR at this resolution, the total number of diffusion directions was increased to 1,000 across all shells (including 3 $\times$ $b=0$, 80 directions each at $b=1$ and $b=2.5$, 128 at $b=5$, 136 at $b=7.35$, and 256 at $b=12.5$ for each $\Delta$), the repetition time needed to be extended to TR = 4100 ms and the matrix size 138 x 138 with 82 axial slices. As a result, the scan duration increased four-fold, and the mean SNR decreased only moderately, from 32 (Confidence Interval (CI): [13, 60]) at 2 mm to 27 (CI: [11, 56]) at 1.6 mm. This 69-minute scan enabled exploratory assessment of spatial fidelity and model robustness at finer anatomical scales.

\subsection{Preprocessing}

For each protocol, all diffusion-weighted volumes were jointly preprocessed using a standardized pipeline. This process began with complex-domain Marchenko-Pastur Principal Component Analysis (MP-PCA) denoising \citep{veraartDenoisingDiffusionMRI2016}. To correct for slowly varying phase variations, a two-pass procedure was employed: an initial MP-PCA pass was run on the complex data, the low-pass filtered phase from this output was used to correct the original data's phase, and a final MP-PCA pass was applied to the phase-corrected data. The real part of this final result was used for subsequent steps. The pipeline then continued with Gibbs ringing correction \citep{kellnerGibbsringingArtifactRemoval2016}, susceptibility-induced distortion and eddy current correction via FSL's \texttt{topup} and \texttt{eddy} tools \citep{anderssonIntegratedApproachCorrection2016}, and finally, gradient nonlinearity correction using scanner-specific maps \citep{huangConnectome20Developing2021}.

$T_1$-weighted images were segmented using FastSurfer \citep{henschelFastSurferFastAccurate2020}, and gray matter regions of interest (ROIs) from the Desikan-Killiany-Tourville (DKT) atlas \citep{klein101LabeledBrain2012} were projected from $T_1$ to diffusion space via affine registration with ANTs \citep{avantsAdvancedNormalizationTools2009}.

\subsection{Model fitting}

For all protocols, powder-averaged diffusion signals were calculated and NEXI parameters were estimated using the same nonlinear least-squares optimization framework \citep{jelescuNeuriteExchangeImaging2022}  to ensure a fair comparison of the information content intrinsic to each acquisition scheme.
Estimated parameters included intra-neurite diffusivity ($D_i$), extra-neurite diffusivity ($D_e$), intra-neurite signal fraction ($f$), and inter-compartment exchange time ($t_{ex}$). Fitting constraints were applied to maintain biological plausibility and fitting stability: parameters were bounded within $[1, 150]$ ms for $t_{ex}$, $[0.1, 3.5]$ µm$^2$/ms for both $D_i$ and $D_e$, and $[0.1, 0.9]$ for $f$. Notably, the upper bound for $D_i$ was chosen slightly above free water diffusivity at body temperature, to leave margin for over-estimation due to CSF partial volume effects (that come with flow/pulsation artifacts) or noise. These constraints were consistent with previous implementations on 3T Prisma and Connectome 1.0 MRI systems \citep{uhlQuantifyingHumanGray2024, uhlHumanGrayMatter2025}. Cortical surfaces and parameter maps were visualized using Connectome Workbench \citep{marcusInformaticsDataMining2011}.
To ensure statistical robustness and avoid kurtosis bias introduced by averaging signals with different underlying microstructural orientations, the model was fitted strictly at the voxel level. Any region-of-interest (ROI) mean signals or aggregated parameter maps presented in this study represent the subsequent averaging of these individual voxel-wise fits, rather than a single fit to a pre-averaged signal.

\subsection{Synthetic simulations and protocol reduction}

A synthetic dataset of $10^6$ diffusion signals was generated with the NEXI model.  Ground-truth parameter vectors were sampled uniformly within empirical bounds obtained from the full 15-feature protocol ($t_{ex}$:[1,70]ms, $f$:[0.15,0.80], $D_i$:[1.7,3.5]ms/µm$^2$, $D_e$:[0.5,1.5]ms/µm$^2$). These empirically derived ranges ensured that the simulations covered only physiologically plausible values while avoiding extreme outliers
observed in low SNR voxels. Gaussian noise was then added so that the resulting SNR matched the in vivo distribution (mean:32, CI:[13, 60]). The four NEXI target parameters were z-scored before training, meaning we subtracted the mean and divided by the standard deviation for each parameter. This step ensures all parameters are on a common scale, which is essential for providing fair, unitless feature attributions. To demonstrate the scalability of our approach beyond the discrete 15-feature master protocol, we also performed an extended optimization on a high-density, physically constrained $(b, \Delta)$ grid, the details of which are provided in the Supplementary Materials (Figure \ref{fig:supp8}).

An XGBoost regression model \citep{chenXGBoostScalableTree2016} was trained to predict each parameter from the full 15-feature input. Feature relevance was quantified using SHAP \citep{lundbergUnifiedApproachInterpreting2017}, and RFE was applied to iteratively remove the least informative features.

The final 8-feature protocol was selected based on: (i) the inflection point in normalized root mean square error (RMSE) across parameters, and (ii) protocol compatibility with time-dependent diffusion kurtosis analysis \citep{fieremansMonteCarloStudy2010,aggarwalDiffusiontimeDependenceDiffusional2020,solomonTimedependentDiffusivityKurtosis2023,novikovTimeDependentDiffusionKurtosis2011} for secondary validation of exchange effects.

\paragraph{Implementation details:}
The XGBoost models were trained with 128 trees, a maximum depth of eight, a learning rate of 0.3, a subsampling ratio of 0.6, and an $\ell_{2}$-regularization coefficient (\texttt{lambda}) of 1.0.  
These hyperparameters were determined through a grid search optimization on a validation set to minimize the mean squared error while preventing overfitting.
Training was stopped early if the validation loss failed to improve for 25 consecutive boosting rounds.  
While the initial one-million synthetic signals were generated uniformly to ensure robust training (split into 80\% training and 10\% validation examples), the remaining 10\% hold-out test examples were resampled to match the realistic in vivo distribution derived from the full protocol results. This ensures that the final performance benchmarking (e.g., Figure \ref{fig:figure3}) reflects biologically plausible conditions rather than a uniform distribution that would over-represent unlikely physiological regimes.
Feature attributions were computed with the tree-SHAP algorithm implemented in the \texttt{shap} Python library \url{https://github.com/shap/shap}. 
Recursive feature elimination proceeded by removing the least important feature at each iteration until only two features remained. 
To ensure reproducibility and provide a benchmark for future users, the pipeline was executed on a laptop equipped with an Intel Core i7-13700H (16 cores), 30 GB of RAM, and an NVIDIA GeForce RTX 4070 Laptop GPU. Using GPU acceleration for XGBoost, the initial training on 1,000,000 samples was completed in 5.05 s, and the entire recursive optimization process (XGBoost-SHAP-RFE) was finalized in 2 minutes and 46 seconds. 
We inspected the resulting normalised RMSE curve and retained the eight feature subset located near its inflection point.  
All scripts, trained models, and the synthetic signal generator are available in the repository cited in the \emph{Data and code availability} section.

\subsection{Comparison of protocol performance}

To benchmark the performance and robustness of the XAI-optimized protocol against both theoretical limits and common expert practices, we evaluated it against the full 15-feature acquisition and three alternative 8-feature subprotocols:

\begin{itemize}
    \item \textbf{Theory-driven Weighted-CRLB:} We implemented a rigorous optimization based on the CRLB. For each simulated ground-truth voxel, the Jacobian matrix was computed and scaled by the corresponding noise standard deviation per shell. To account for the uneven number of directions per shell, the noise variance term in the CRLB calculation was appropriately scaled for each shell as $\sigma^2_{\text{shell}} = \sigma^2_{\text{voxel}} / N_{\text{dir}}$, ensuring that shells with more directions were correctly weighted as having higher precision. The inverse of the resulting FIM provided the local CRLB covariance matrix. To avoid bias from the different orders of magnitude and numerical ranges of the NEXI parameters (e.g., $t_{ex} \in [1, 150]$ ms vs $f \in [0.1, 0.9]$), we used a weighted A-optimality criterion. More precisely, the diagonal elements of the CRLB matrix, representing parameter variances, were normalized by the square of their respective ground-truth parameters, converting them into squared coefficients of variation. We then minimized the trace of this weighted covariance matrix, averaged across the entire physiological prior distribution. This protocol represents the theoretical gold standard under Gaussian noise assumptions.
    
    \item \textbf{Heuristic "Mid-Range":} This protocol mimics a standard multi-shell acquisition compatible with models like SANDI but with an additional $\Delta$. It densely samples b-values at a single intermediate diffusion time ($\Delta = 27$ ms), supplemented by limited corner points ($b=1, 2, 12.5$ ms/µm$^2$) at $\Delta = 45$ ms. This design prioritizes b-value coverage but lacks temporal diversity.
    
    \item \textbf{Heuristic "Corner":} Responding to the design principle of maximizing the lever arm, this protocol samples only the extremes of the acquisition space: the lowest and highest b-values at the shortest and longest $\Delta$. While theoretically maximizing sensitivity to signal decay, this approach is potentially vulnerable to noise at high b-values.
\end{itemize}

The comparison across all protocols was multifaceted. In simulations, we assessed the absolute estimation error. For the in vivo data, comparison metrics included the root mean squared deviation (RMSD) of region-of-interest (ROI) medians between protocols, the similarity of parameter distributions across the cortex, and the visual agreement of the resulting cortical surface maps.

Test-retest repeatability was assessed by computing Bland-Altman plots of ROI medians across repeated sessions for each protocol. Limits of agreement and estimation bias were calculated separately for all four NEXI parameters.

% ========= Results =========

\section{Results}

\subsection{Cortical mapping with the full protocol}
\begin{figure}[H]
\centering
\includegraphics[width=\textwidth]{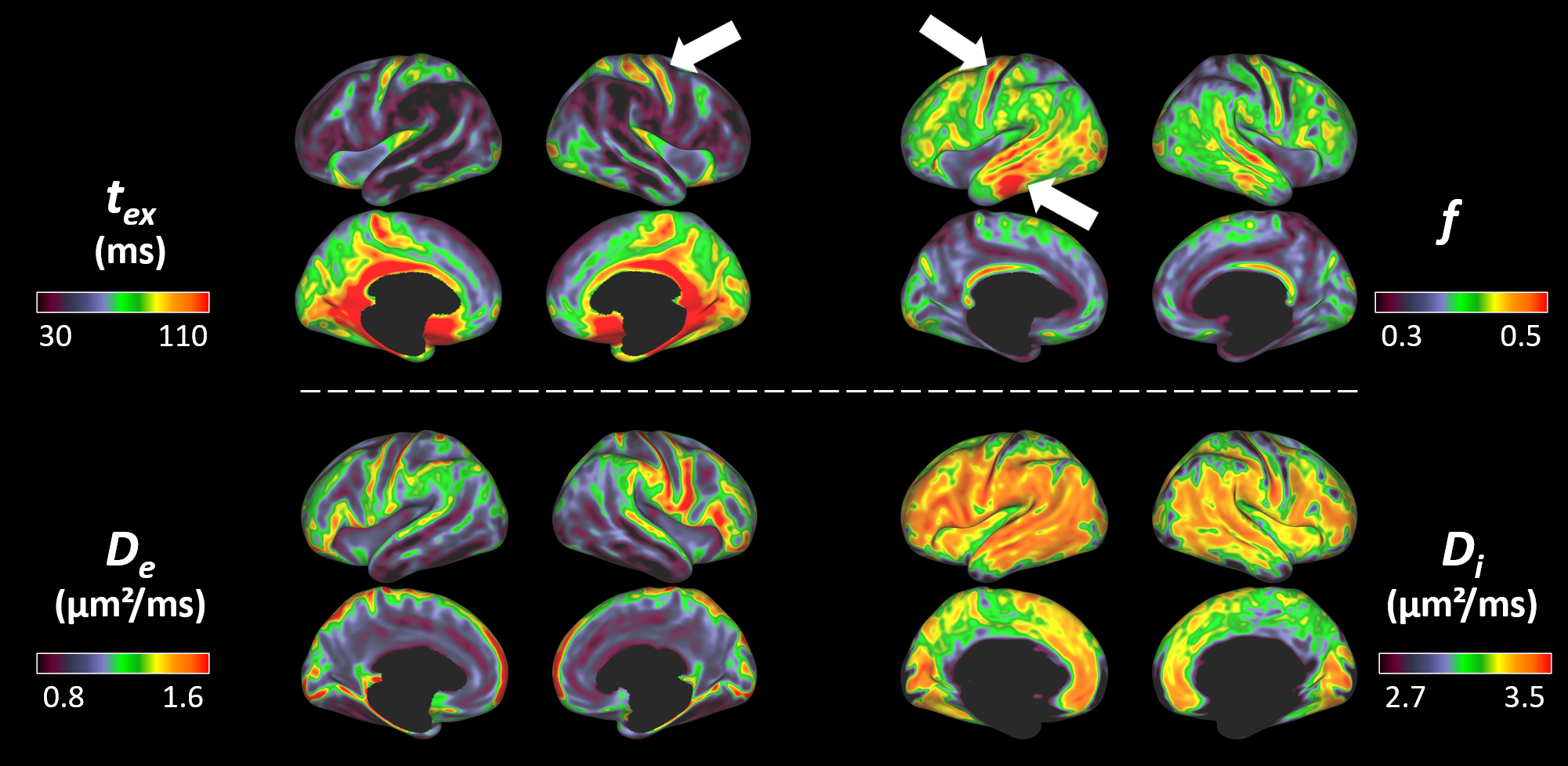}
\caption{
\textbf{Cortical surface maps of NEXI parameters from the full protocol.}  
Group-averaged maps of $t_{ex}$, $f$, $D_i$, and $D_e$ across all subjects, displayed on lateral and medial views of both cortical hemispheres. As expected, $t_{ex}$ and $f$ are elevated in the motor cortex, consistent with prior associations to increased myelination and neurite density. However, unlike earlier reports on Connectome 1.0 and Prisma systems, no clear elevation of $t_{ex}$ is observed in the visual cortex or temporal lobe. Interestingly, $f$ is elevated instead in the temporal lobe. In contrast, $D_i$ remains relatively uniform across the cortex ($3$ µm$^2$/ms), while $D_e$ appears reduced in areas with high $f$, potentially reflecting more restricted extracellular environments. These full-protocol maps are used as the spatial reference for evaluating the fidelity of reduced protocols.
}
\label{fig:figure1}
\end{figure}

Figure \ref{fig:figure1} shows group-averaged cortical maps of NEXI parameters estimated from the full 15-feature protocol. As previously reported \citep{uhlQuantifyingHumanGray2024, uhlHumanGrayMatter2025, chanVivoHumanNeurite2025}, elevated values of $t_{ex}$ and $f$ are observed in the sensorimotor cortex, suggesting increased neurite packing and myelination. Interestingly, high $t_{ex}$ values are not observed in visual or temporal regions, unlike prior studies on Connectome 1.0 and Prisma. Our finding for the visual cortex also contrasts with a previous report on the Connectome 2.0 system \citep{chanVivoHumanNeurite2025}. Instead, high $f$ values are estimated in the temporal lobe. The intra-neurite diffusivity $D_i$ remains largely homogeneous across the brain, often hitting the enforced upper bound ($3.5$ µm$^2$/ms), while $D_e$ is slower in high-$f$ regions, consistent with reduced extracellular space.

% ---------- Figure 1 ----------

\subsection{Data-driven protocol reduction using XAI}
\begin{figure}[H]
\centering
\includegraphics[width=\textwidth, clip=false]{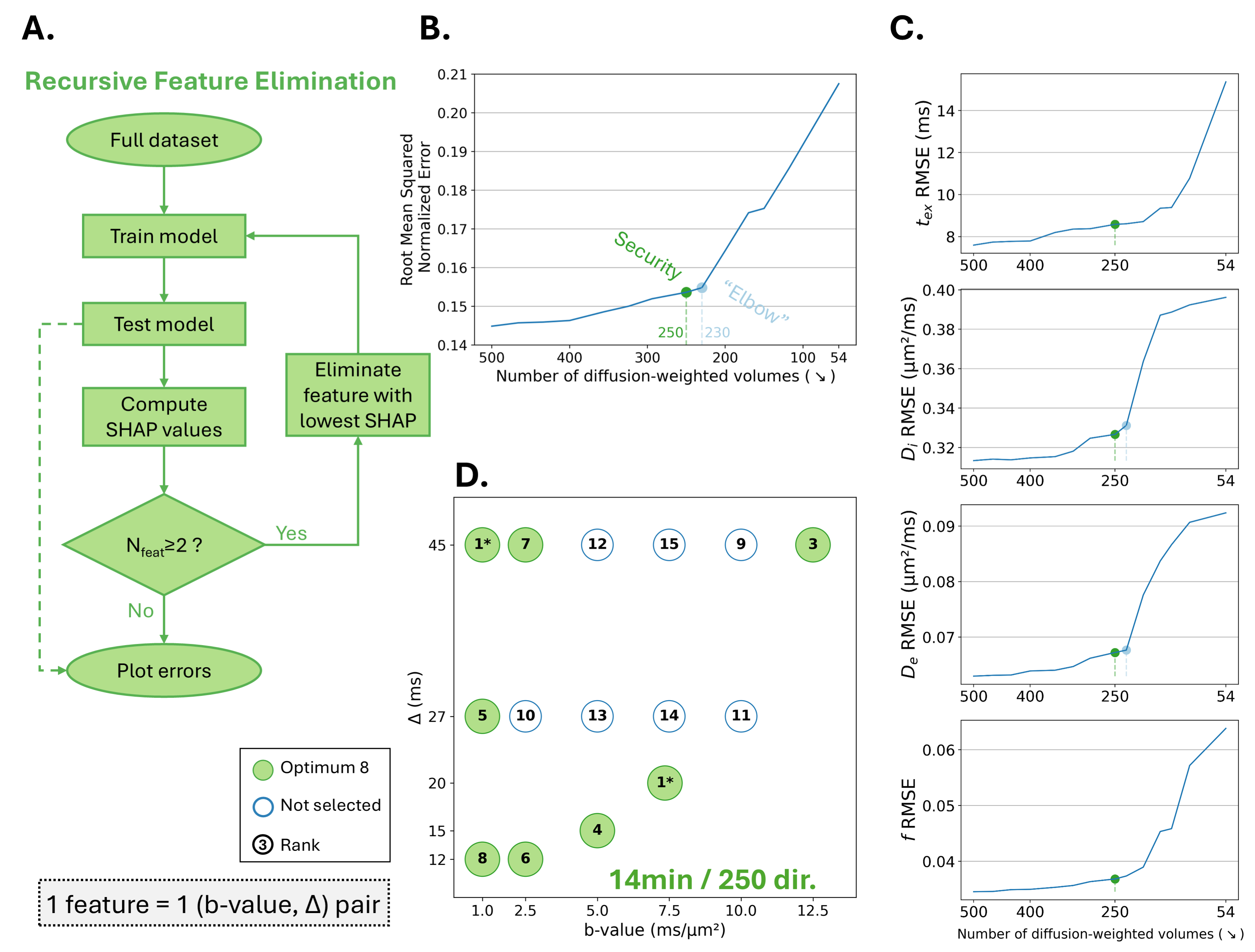}
\caption{
\textbf{Recursive feature elimination using SHAP for NEXI protocol optimization.}
(A.) Overview of the XAI-RFE pipeline: simulated NEXI signals were used to train XGBoost regressors, and SHAP values quantified each feature's contribution to parameter estimation. Features (defined by unique $b$ and $\Delta$ values) were iteratively removed based on minimal SHAP importance. Evolution of normalized RMSE across all (B.) and each (C.) parameters as features (and thus diffusion-weighted volumes) are removed. An inflection at 7-8 features (totalizing 230-250 diffusion-weighted volumes across the remaining features) indicates the minimal protocol length before accuracy sharply degrades. (D.) Visualization of the final selected protocol (green), overlaid on the full 15-feature set. The retained features span both short and long diffusion times and a broad range of b-values.
}
\label{fig:figure2}
\end{figure}

To identify a reduced acquisition protocol, we applied a machine learning-based recursive feature elimination strategy using SHAP values computed from XGBoost regressors trained on synthetic NEXI signals. Figure \ref{fig:figure2} outlines this process. It revealed an inflection point at 7 features in the root mean squared error (RMSE) curves, beyond which additional features yielded marginal improvements (Fig. \ref{fig:figure2}). The primary motivation for retaining this specific additional acquisition (b=$1$ms/µm$^2$ at a short $\Delta$) is that it allows for the estimation of diffusion kurtosis at a second diffusion time, thereby enabling time-dependent DKI analysis \citep{fieremansMonteCarloStudy2010, novikovTimeDependentDiffusionKurtosis2011} and a secondary measurement of $t_{ex}$ beyond the NEXI model itself. To rigorously assess the stability of this feature ranking, we repeated the entire XAI-RFE optimization process using 15 different random seeds. In every instance, the pipeline converged to the exact same 8-feature protocol. This confirms that the feature selection is highly robust to initialization variability and reflects the underlying physical geometry of the signal rather than stochastic noise in the training process. The retained features covered a broad range of b-values (1.0-12.5 ms/µm$^2$) and diffusion times ($\Delta$ from 15-45 ms), ensuring sensitivity to both restricted diffusion and exchange effects.

% ---------- Figure 2 ----------

\subsection{Validation on synthetic signals and in vivo distributions}
\begin{figure}[H]
\centering
\includegraphics[width=0.80\textwidth]{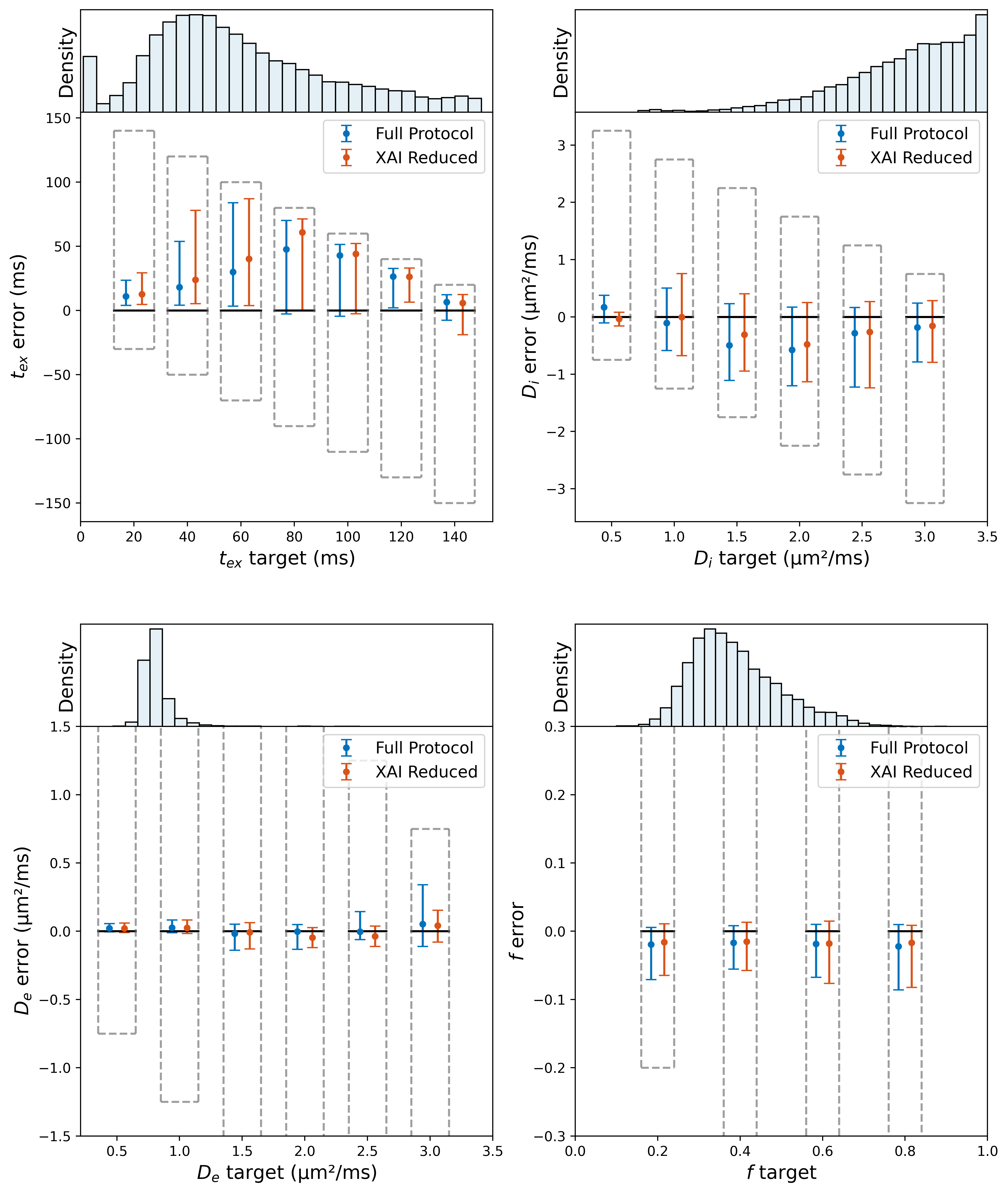}
\caption{
\textbf{Estimation error across the parameter space: Full vs. XAI-optimized protocol.}
Binned error analysis from 100{,}000 synthetic NEXI signals with added Gaussian noise (SNR = 32 ± 26), comparing the full protocol (blue) and the reduced 8-feature XAI protocol (orange). Parameters were estimated using a classical non-linear least squares (NLS) solver for both the full (blue) and reduced (orange) protocols. For each NEXI parameter ($t_{ex}$, $f$, $D_i$, $D_e$), absolute estimation errors are plotted as a function of the ground-truth parameter value. Note that y-axis limits are scaled specifically for each parameter to optimize the visibility of smaller error margins. Histograms above each panel indicate the distribution of target values in the test set. Error bars represent mean ± standard deviation within each bin; dashed rectangles denote model fitting bounds.
Both protocols show similar error profiles across the full physiological range. A systematic positive bias in $t_{ex}$ is observed, reaching 20-40 ms for intermediate values, likely due to model limitations at short diffusion times or imperfect sensitivity to fast exchange. Estimates of $D_i$ tend to be underestimated at intermediate values. Errors for $f$ and $D_e$ remain low and stable. These results confirm that the XAI-reduced protocol preserves parameter accuracy and robustness under realistic conditions.
}
\label{fig:figure3}
\end{figure}

Validation on 100{,}000 synthetic noisy signals (SNR = 32 ± 26 at b=0) (hold-out set, not used during any stage of the XGBoost training or RFE process) demonstrated that the XAI-reduced protocol preserves estimation accuracy across the full physiological range of NEXI parameters (Figure \ref{fig:figure3}). A consistent overestimation of $t_{ex}$ is observed for intermediate values, likely related to limited sensitivity at short $\Delta$. $D_i$ tends to be underestimated at intermediate values. Despite this, the reduced protocol shows estimation profiles nearly identical to the full acquisition. Paradoxically, the reduced protocol can even marginally outperform the full protocol in specific error bins (e.g., for $f$). This slight improvement occurs because discarding certain high b-value shells that are heavily dominated by noise effectively reduces the fitting instability for these specific parameters. Crucially, parameter estimation on this hold-out set was conducted using a standard non-linear least squares (NLS) fitting procedure rather than the XGBoost model used for optimization. This ensures that the identified 8-feature protocol is intrinsically informative and that its performance is not dependent on a specific machine-learning estimator.

% ---------- Figure 3 ----------

\begin{figure}[H]
\centering
\includegraphics[width=\textwidth]{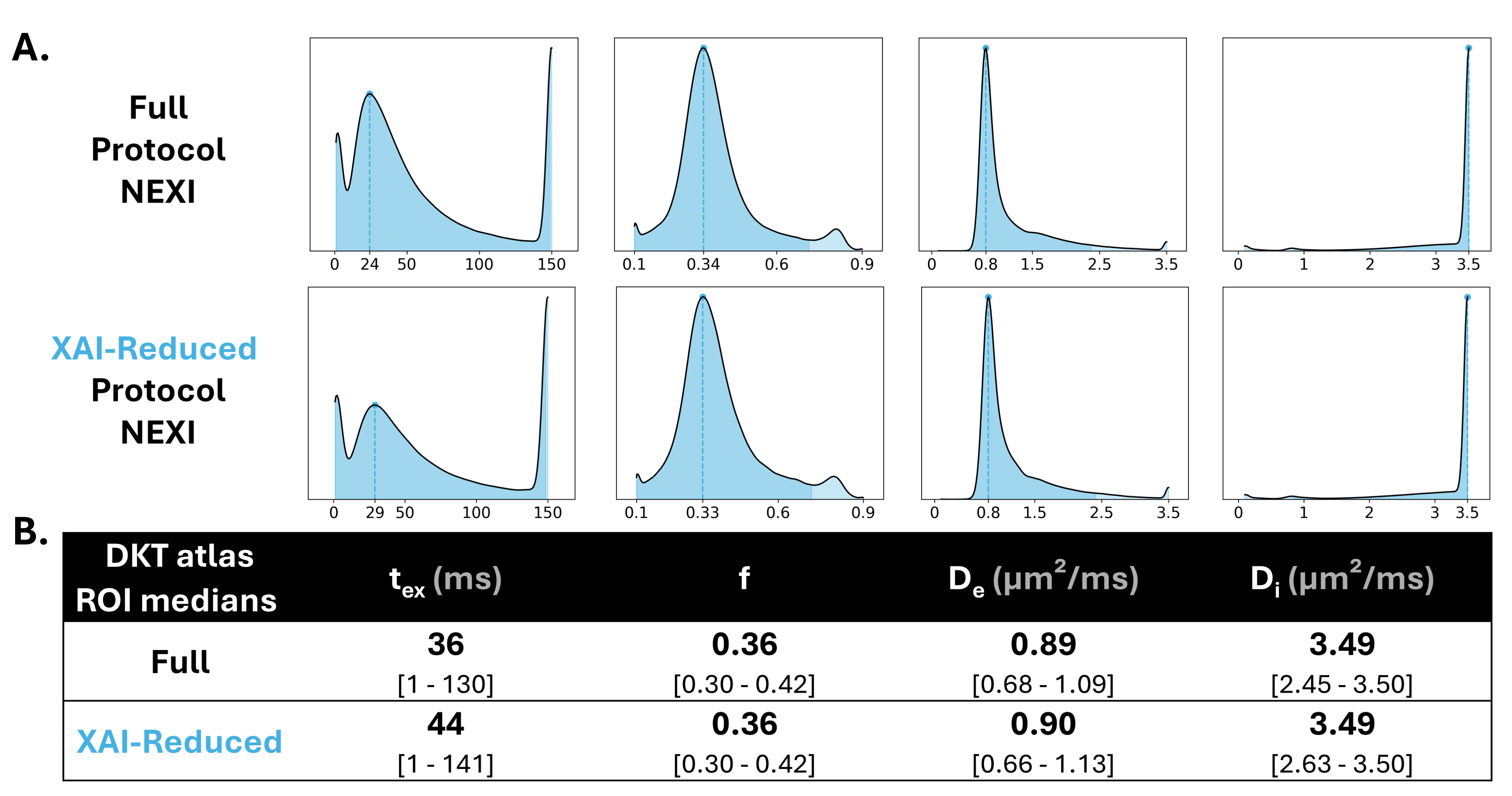}
\caption{
\textbf{Distributions and summary of NEXI parameter estimates across DKT cortical regions.}
(A) Kernel density estimates of voxel-level values for each NEXI parameter ($t_{ex}$, $f$, $D_e$, $D_i$), aggregated across all DKT ROIs and all subjects. The full protocol (top row) and XAI-reduced protocol (bottom row) yield very similar distributions. The vertical dashed lines indicate the mode of each distribution, reflecting the most frequent voxel value. Both protocols exhibit similar features: a strong ceiling effect for $D_i$ at the upper model bound ($3.5$ µm$^2$/ms), right-skewed $t_{ex}$ distributions with local peaks around 25-30 ms, and unimodal distributions for $f$ and $D_e$. 
(B) Summary statistics computed from ROI-wise medians across the cortex. Median values and ranges are consistent between protocols, confirming the robustness of the XAI-reduced protocol.
}
\label{fig:figure4}
\end{figure}

Voxel-level distributions and regional medians in healthy volunteers confirm this consistency (Figure \ref{fig:figure4}). Although the reduced protocol yields a slightly heavier tail in the $t_{ex}$ distribution (as a trade-off to reducing the datapoints by 50\%), modes and median values across DKT cortical ROIs show minimal differences between protocols, with $t_{ex}$ around 40 ms, $f \approx 0.36$, and $D_e \approx 0.9$ µm$^2$/ms. These statistics (mode and median) are a priority fot group-level analyses. These results therefore support the robustness of the XAI-reduced protocol across typical inter-subject variability.

% ---------- Figure 4 ----------

\subsection{Benchmark against theory-driven and heuristic protocols}

\begin{figure}[H]
\centering
\includegraphics[width=\textwidth]{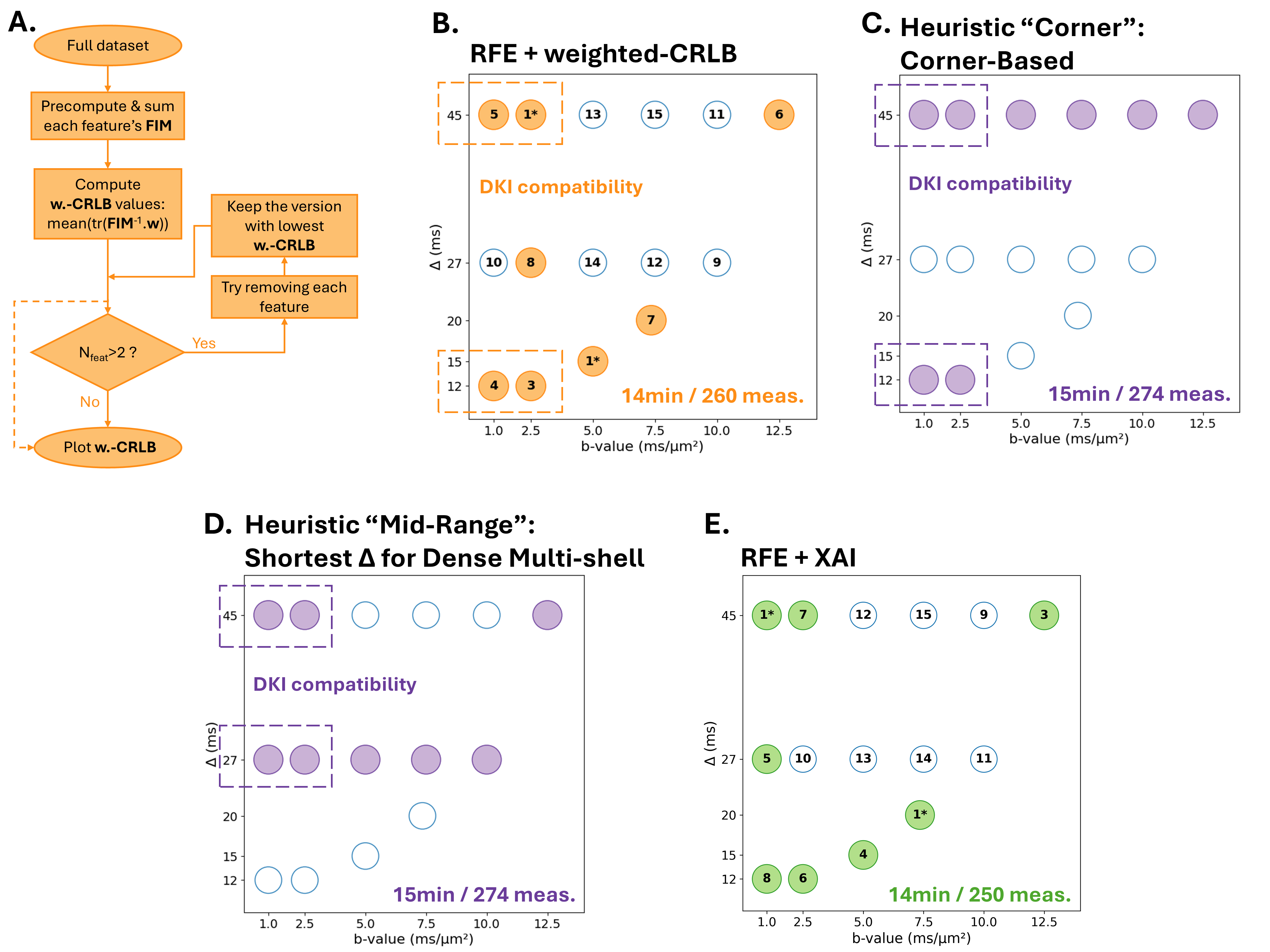}
\caption{
\textbf{Design of theory-driven and heuristic comparison protocols.}
(A) Schematic of the Weighted CRLB-based Recursive Feature Elimination algorithm. The process iteratively removes features to minimize the mean relative variance (A-optimality) across a range of ground truths.
(B) The resulting \textbf{Weighted CRLB protocol} is strikingly similar to the XAI-derived protocol, selecting features at the boundaries of the parameter space while maintaining temporal diversity.
(C) The \textbf{Heuristic "Corner"} protocol samples only the extremes of the ($b$, $\Delta$) space to maximize the geometric lever arm.
(D) The \textbf{Heuristic "Mid-Range"} protocol densely samples a single intermediate diffusion time ($\Delta=27$ ms); this specific subset forms a standard multi-shell scheme compatible with fixed-diffusion-time models like SANDI. The protocol adds only limited coverage at $\Delta=45$ ms, resulting in an overall lack of temporal spread critical for robust exchange mapping.
(E) The \textbf{XAI-optimized protocol} (repeated for comparison) balances these constraints, achieving a configuration that mirrors the theoretical optimum (CRLB) while being derived purely from data-driven feature importance.
}
\label{fig:figure5}
\end{figure}

To rigorously assess the optimality of the XAI-derived protocol, we benchmarked it against three alternative 8-feature subprotocols: a theoretically optimal protocol derived from the CRLB, and two expert-designed heuristic protocols representing common acquisition strategies (Figure \ref{fig:figure5}).

The theory-driven protocol was obtained using a weighted A-optimality criterion based on the CRLB. This method minimizes the sum of the relative variances (squared coefficients of variation) of the estimated parameters, averaged across a broad range of physiological ground truths. Remarkably, this purely theoretical optimization converged to a protocol nearly identical to the one identified by our data-driven XAI approach (Fig. \ref{fig:figure5}B vs. E), providing strong independent validation of the XAI selection. Furthermore, to ensure these findings were not artifacts of the discrete 15-feature master protocol, both XAI and CRLB optimizations were extended to a high-density acquisition grid (Supplementary Figure \ref{fig:supp8}). Both methods consistently converged toward the same macroscopic strategy, confirming the robust nature of the derived protocol.

The two heuristic protocols highlight the pitfalls of intuitive design. The "Heuristic Corner" protocol (Fig. \ref{fig:figure5}C) samples the extremes of the acquisition space to maximize the geometric lever arm. While theoretically sound for noiseless data, this approach proved highly unstable in practice. The "Heuristic Mid-Range" protocol (Fig. \ref{fig:figure5}D) mimics a standard multi-shell acquisition with dense sampling at intermediate diffusion times ($\Delta=27$ ms), compatible with models like SANDI, but lacks the temporal diversity required for robust exchange mapping.

\begin{figure}[H]
\centering
\includegraphics[width=\textwidth]{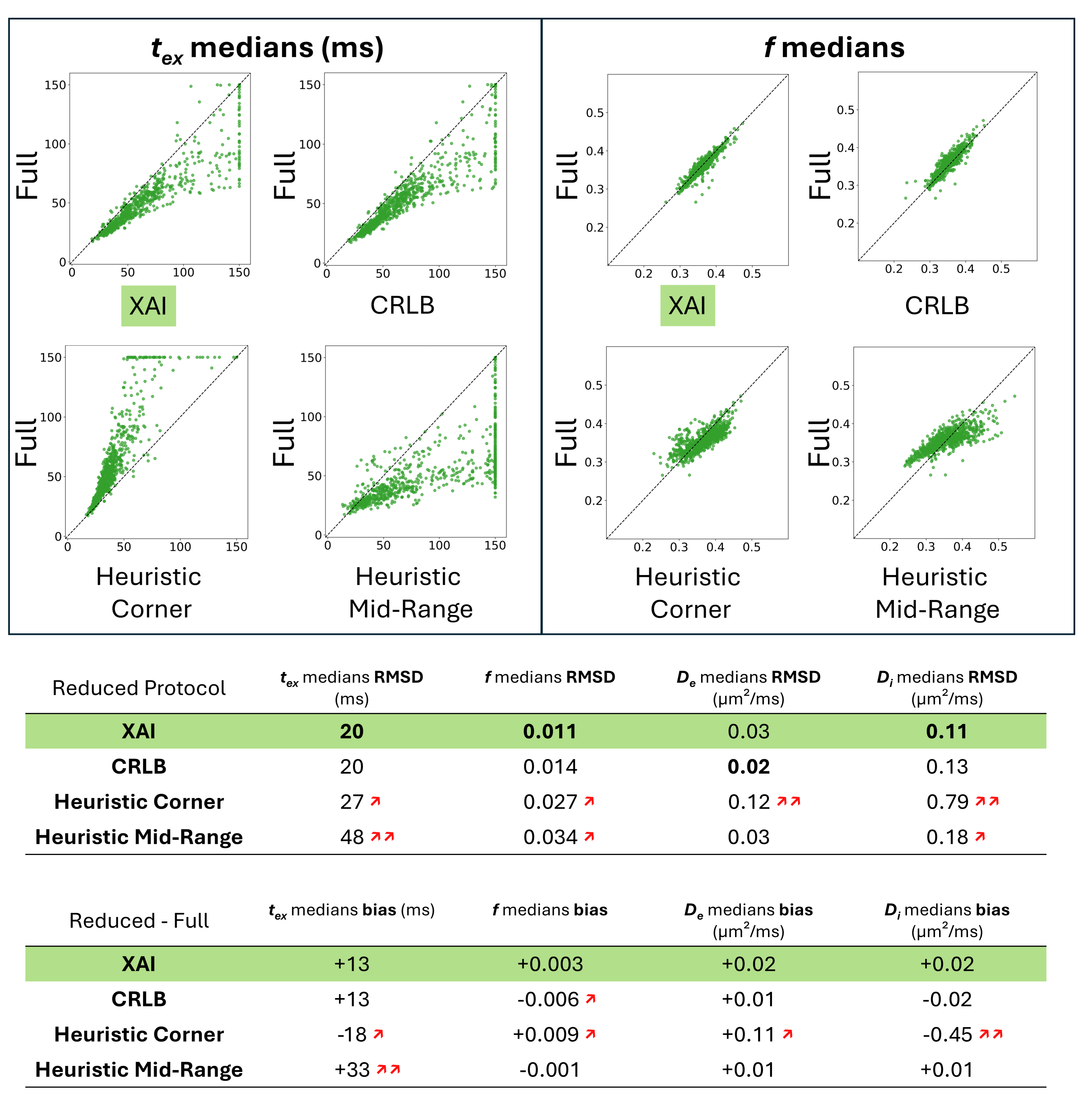}
\caption{
\textbf{Quantitative benchmark: RMSD and estimation bias across protocols.}
Top: Scatter plots comparing DKT ROI medians from the full protocol (y-axis) against each reduced protocol (x-axis) for $t_{ex}$ and $f$, aggregated across all ROIs and subjects. The XAI and CRLB protocols show good agreement with the full acquisition (identity line). The "Heuristic Corner" protocol shows increased dispersion, while the "Heuristic Mid-Range" protocol exhibits a systematic bias for $t_{ex}$.
Bottom: Table summarizing the Root Mean Square Deviation (RMSD) and mean bias for all parameter ROI medians. Values in bold indicate the best performance. Red arrows highlight significant performance degradation compared to the XAI protocol. The XAI protocol achieves the lowest RMSD and bias overall. Notably, the "Heuristic Mid-Range" protocol shows a large bias in $t_{ex}$ (+33 ms) due to limited $\Delta$ diversity, while the "Heuristic Corner" protocol suffers from severe instability in $D_i$ (RMSD 0.79) and negative bias (-0.45), confirming its susceptibility to noise.
}
\label{fig:figure6}
\end{figure}

Quantitative comparisons of in vivo estimations (Figure \ref{fig:figure6}) reveal that the XAI-optimized protocol consistently outperforms the heuristic alternatives. It yields the lowest Root Mean Square Deviation (RMSD) from the full protocol across all parameters. The "Heuristic Mid-Range" protocol exhibits a substantial bias in $t_{ex}$ (+33 ms) and a 2.4-fold higher RMSD compared to XAI, confirming that insufficient $\Delta$ sampling severely compromises exchange time estimation. Conversely, the "Heuristic Corner" protocol suffers from massive instability in $D_i$ (RMSD 0.79 vs 0.11 $\mu m^2/ms$ for XAI) and a significant negative bias (-0.45 $\mu m^2/ms$), demonstrating that extreme sampling is vulnerable to noise at high b-values. The weighted CRLB protocol performs comparably to XAI, though with slightly higher deviation in $f$ and $D_i$.

% ---------- Figure 5 ----------

% ---------- Figure 6 ----------

\begin{figure}[H]
\centering
\includegraphics[width=\textwidth]{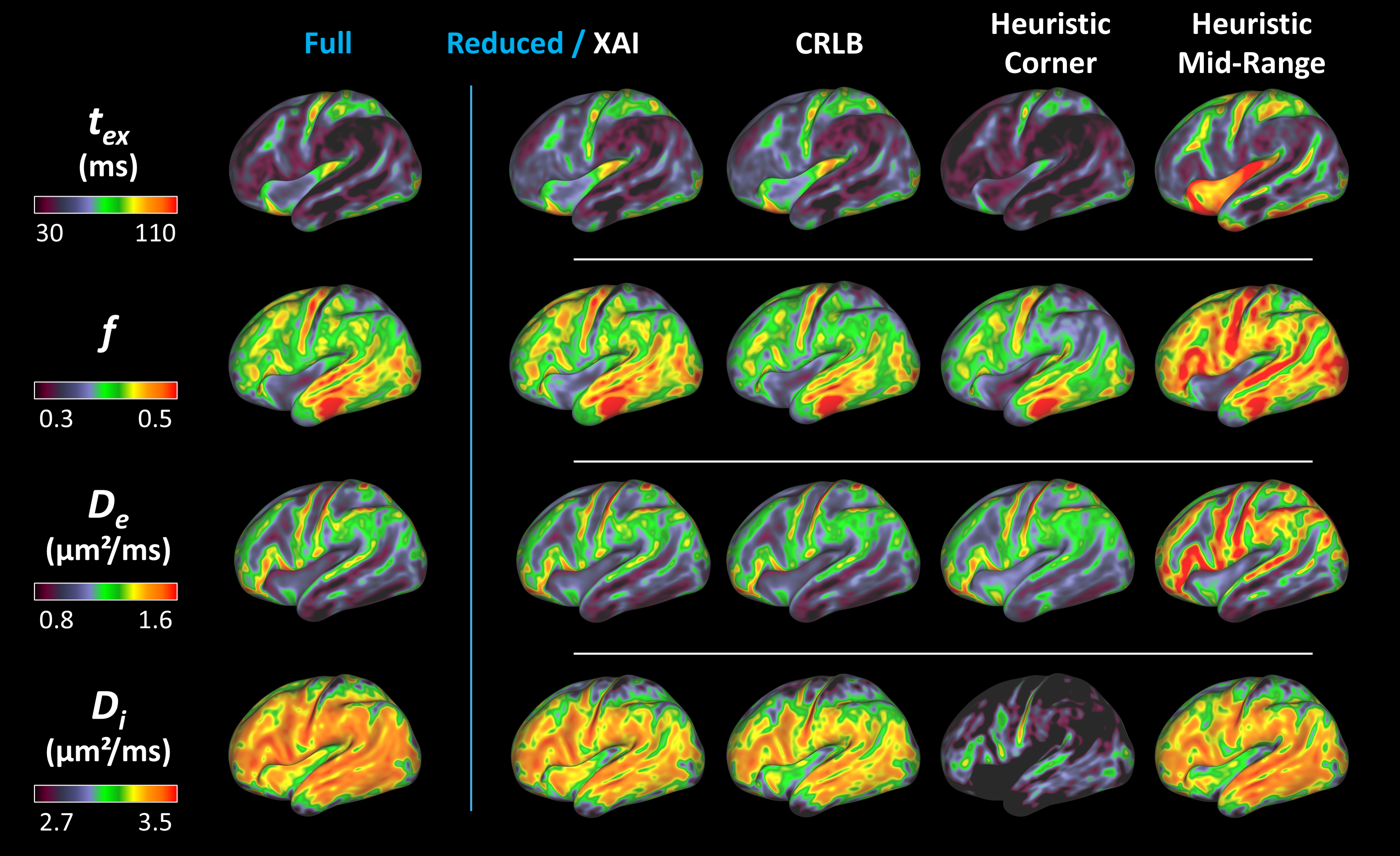}
\caption{
\textbf{Cortical surface maps of NEXI parameters comparisons across protocols.}
Group-averaged maps for $t_{ex}$, $f$, $D_e$, and $D_i$ are displayed for the Full protocol, the XAI-optimized reduced protocol, and the three benchmark protocols (CRLB, Heuristic Corner, Heuristic Mid-Range).
The XAI protocol (second column) is visually indistinguishable from the Full protocol (first column), preserving all key anatomical contrasts.
The \textbf{CRLB protocol} also performs well but shows slightly more variance in $f$.
The \textbf{Heuristic Corner protocol} fails catastrophically for $D_i$ (bottom row), producing a dark map indicative of underestimation due to noise sensitivity at high b-values.
The \textbf{Heuristic Mid-Range protocol} yields a distorted $t_{ex}$ map (top row) with exaggerated values in the insular cortex and an increased contrast in the temporal lobe, confirming that insufficient diffusion time sampling biases exchange estimates.
}
\label{fig:figure7}
\end{figure}

Surface parameter maps (Figure \ref{fig:figure7}) visually reinforce these quantitative findings. The XAI protocol faithfully reproduces the spatial contrasts of the full protocol, preserving the elevated $t_{ex}$ and $f$ patterns in the sensorimotor and visual cortices. In contrast, the "Heuristic Mid-Range" protocol yields a flattened, over-smoothed $t_{ex}$ map that fails to capture regional variation, while the "Heuristic Corner" protocol produces a noisy and artifact-prone $D_i$ map, with extensive areas hitting the lower estimation bounds (dark regions). These results underscore that neither maximizing b-value range alone nor maximizing sampling density at a single $\Delta$ is sufficient; the XAI-identified sampling strategy provides the optimal compromise for robust microstructural mapping.

% ---------- Figure 7 ----------

\subsection{Test-retest reliability and high-resolution scalability}
\begin{figure}[H]
\centering
\includegraphics[width=0.7\textwidth]{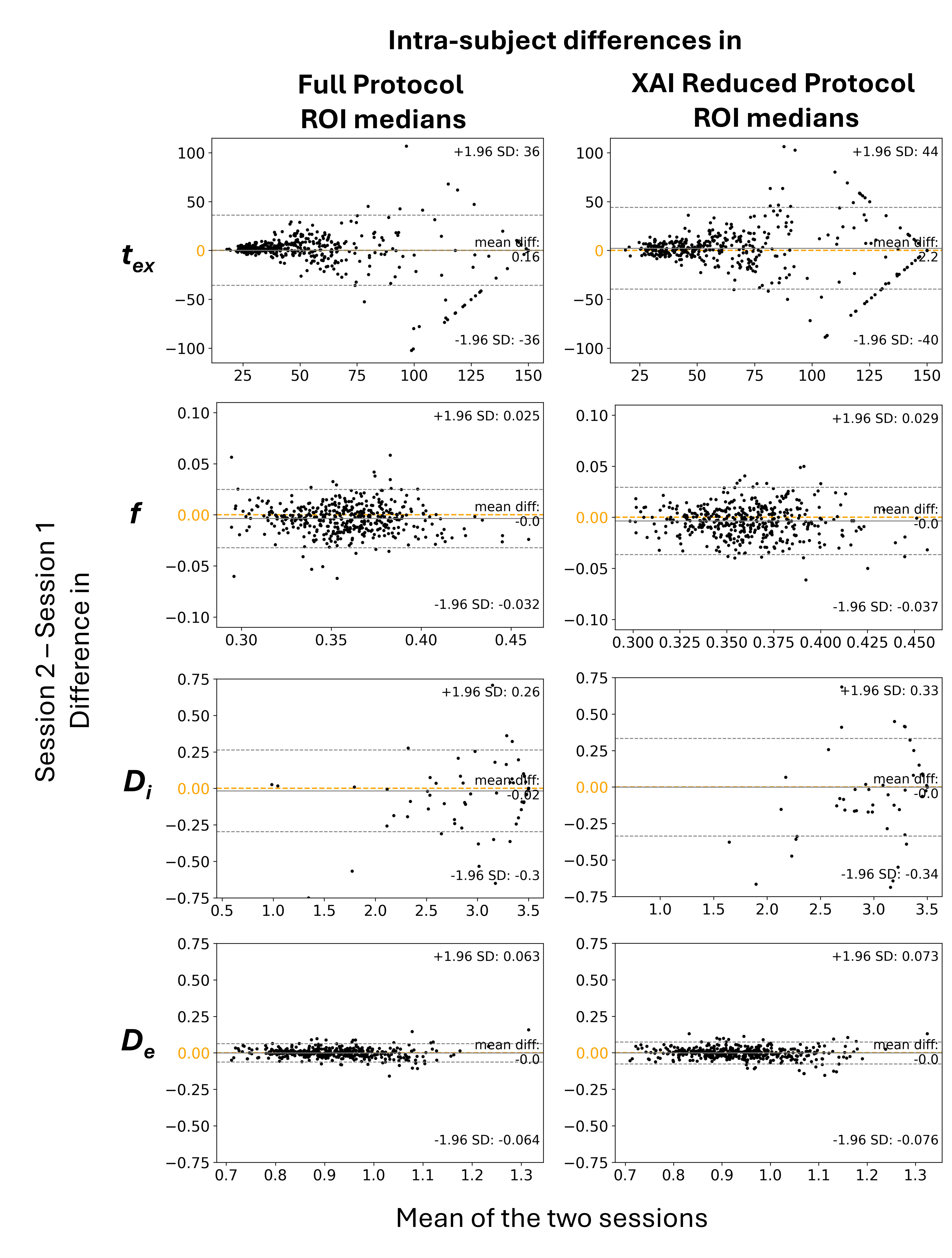}
\caption{
\textbf{Test-retest reproducibility of NEXI estimates using the full protocol and the XAI-optimized subprotocol.}
Bland-Altman plots show intra-subject variability across scan-rescan sessions, using DKT ROI median values for each NEXI parameter ($t_{ex}$, $f$, $D_i$, $D_e$). 
Left: Full protocol. Right: XAI-Reduced protocol. 
Each plot displays the difference between sessions against their mean, with orange lines indicating bias and dotted gray lines denoting the 95\% limits of agreement. 
The XAI protocol maintains reproducibility comparable to the full protocol, with marginally wider limits for $t_{ex}$ and $D_i$ but near-identical performance for $f$ and $D_e$. 
This supports the use of the 8-feature reduced protocol in longitudinal studies where acquisition time is constrained.
}
\label{fig:figure8}
\end{figure}

Test-retest reproducibility was preserved with the reduced protocol. Bland-Altman analysis (Figure \ref{fig:figure8}) shows that the 8-feature protocol maintains agreement comparable to the full protocol across all parameters, with only marginal increases in the limits of agreement. The seemingly sparse distribution of data points in the $D_i$ Bland-Altman plots is a direct consequence of a strong ceiling effect: a large proportion of voxels consistently hit the upper model bound ($3.5$ \textmu m$^2$/ms) in both scan sessions, causing their coordinates to perfectly overlap on the limit line. In comparison, the alternative reduced protocols (See Figure \ref{fig:supp3}) exhibited distinct trade-offs. While the Weighted CRLB protocol showed reproducibility profiles similar to the XAI protocol, the Heuristic Corner protocol demonstrated severe instability for $D_i$. Conversely, the Heuristic Mid-Range protocol, though reproducible, yields the biased estimates detailed previously. This underscores that the XAI-optimized approach achieves the most robust balance between parameter stability and sensitivity across the full model space.

% ---------- Figure 8 ----------

\medskip

To quantify stability, we computed the mean within-subject Coefficient of Variation (wsCV) for each parameter across all cortical ROIs (Table \ref{tab:reproducibility_mean_cv}). The XAI protocol demonstrates excellent stability, nearly matching the full protocol's performance. In stark contrast, the alternative reduced protocols exhibited distinct failure modes. While the Heuristic Mid-Range protocol yielded reproducible but biased estimates (as seen in Fig. \ref{fig:figure6}), the Heuristic Corner protocol exhibited a polarized behavior. Paradoxically, it achieved the lowest wsCV for $t_{ex}$ (8.4\%). While the XAI and CRLB protocols also favor the boundaries of the $(b, \Delta)$ parameter space, they critically maintain temporal diversity by retaining intermediate diffusion times ($\Delta = 15, 20, 27$ ms). In contrast, the Heuristic Corner protocol allocates its entire budget to the absolute extremes of $\Delta$, heavily oversampling the longest diffusion time ($\Delta = 45$ ms) while ignoring intermediate steps. This massive polarization maximizes the geometric lever arm for the exchange rate, artificially locking in the $t_{ex}$ estimate and reducing its variance. However, it lacks the intermediate temporal support needed to robustly capture the signal decay against noise at high b-values. Consequently, the wsCV for $D_i$ spiked to 5.7\% in the Corner protocol, a nearly 5-fold increase compared to the XAI-optimized protocol (1.3\%) and the full acquisition (1.2\%). This demonstrates that both data-driven and theory-based optimization effectively balance the trade-off between maximizing temporal leverage and preserving full-model stability, a balance missed by extreme heuristic sampling.

% ---------- Table 2 ----------
\begin{table}[h!]
\centering
\setlength{\fboxsep}{10pt}
\fbox{%
\begin{tabular}{@{}lcccc@{}}
\toprule
Protocol & $t_{\text{ex}}$ (\%) & $D_i$ (\%) & $D_e$ (\%) & $f$ (\%) \\
\midrule
Full Protocol       & 10.9 & 1.2 & 1.8 & 2.2 \\
\addlinespace
\textbf{XAI-Reduced} & 12.8 & 1.3 & 2.1 & \textbf{2.5} \\
\addlinespace
CRLB-Reduced                  & 12.4 & 2.0 & \textbf{1.9} & 2.6 \\
Heuristic Corner                & \textbf{8.4}  & 5.7 & 2.0 & 2.6 \\
Heuristic Mid-Range             & 12.2 & \textbf{1.2} & \textbf{1.9} & 2.9 \\
\bottomrule
\end{tabular}
}
\caption{\textbf{Quantification of test-retest reproducibility across acquisition protocols.} 
Values represent the mean within-subject Coefficient of Variation (wsCV, in \%) for each NEXI parameter, calculated across all subjects and cortical ROIs. 
The wsCV is defined as the standard deviation of paired scan-rescan measurements divided by their mean. 
We report the mean wsCV here specifically to expose the instability of the intra-neurite diffusivity ($D_i$) in the \textit{Heuristic Corner} protocol ($5.7\%$) compared to the Full protocol ($1.2\%$). This instability, driven by noise sensitivity at high b-values, is masked when looking solely at median statistics (which are near $0\%$ due to boundary constraints).}
\label{tab:reproducibility_mean_cv}
\end{table}

Beyond protocol reduction, we explored the performance of the XAI-optimized protocol in higher-resolution settings. A 69-minute scan with 1.6 mm isotropic voxels was acquired in one subject, with an increased number of directions to retain the SNR. As shown in Figures \ref{fig:supp5} and \ref{fig:supp6}, spatial distributions of NEXI parameters remained qualitatively consistent with those at 2 mm, with enhanced delineation of the cortical ribbon and sharper gray–white matter boundaries. These findings indicate that the reduced protocol maintains pattern stability under increased spatial resolution, helping to mitigate partial volume effects in cortical gray matter.

We also evaluated the impact of scanner choice and denoising strategy. Cross-scanner comparisons (Tables \ref{tab:suppt1} and \ref{tab:suppt2}) revealed consistent $f$ and $D_e$ estimates between systems, but longer $t_{ex}$ and elevated $D_i$ values on Connectome 2.0, likely due to increased sensitivity to fast intra-neurite dynamics enabled by higher gradient strength \citep{huangConnectome20Developing2021}. However, $D_i$ frequently reaches the model's upper bound, raising the possibility of unmodeled soma effects or fit instabilities.

Finally, we assessed the influence of denoising. Although complex MP-PCA denoising offers, in principle, a way to avoid Rician floor effects that confound signal interpretation at high b-values, magnitude-based processing with Rician correction currently yielded better model fits (Figure \ref{fig:supp4}). This highlights the need for further refinement of complex-domain denoising methods tailored to high-gradient acquisitions.

% ========= Discussion =========
\section{Discussion}

A fundamental barrier hinders the broader adoption of advanced diffusion models: the tension between their biophysical richness and the pragmatic need for time-efficient acquisitions. Optimizing the acquisition time would help promote wider adoption of NEXI in research and clinical applications. For this purpose, we leveraged a state-of-the-art research scanner, the Connectome 2.0, as it offers a unique capability to access a broader range of acquisition parameters enabled by its ultra-high gradient performance. This study successfully demonstrates the feasibility of robustly estimating Neurite Exchange Imaging (NEXI) parameters in human cortical gray matter using a substantially reduced diffusion MRI protocol consisting of only eight strategically selected ($b$, $\Delta$) combinations. Leveraging a data-driven optimization framework based on explainable machine learning (XGBoost-SHAP-RFE or XAI), we achieved a two-fold reduction in scan time on the Connectome 2.0 platform, from 27 to 14 minutes, while preserving key features of spatial contrast, parameter accuracy, and test-retest reproducibility. These findings provide a critical advancement toward the clinical and translational implementation of exchange-sensitive diffusion models, particularly in cortical regions where scan time constraints, noise susceptibility, and parameter degeneracy have traditionally hindered adoption.

\subsection{Advantages of SHAP-guided empirical protocol reduction}

The primary methodological contribution of this study lies in the use of SHAP-guided RFE to empirically optimize a heuristically designed diffusion protocol. Unlike conventional approaches based on the CRLB or the FIM \citep{jalnefjordOptimizationBvalueSchemes2019, flahertyRobustDesignBiological2005}, our method operates without relying on a purely theoretical framework. While the CRLB provides a rigorous lower bound on the local variance of an unbiased estimator, it struggles to account for physical boundary constraints (e.g., the strict upper bound on $D_i$) and the non-linear estimation biases inherent to the fitting landscape. Even when extended to Rician noise \citep{karlsenParameterEstimationRiciandistributed1999}, analytical optimization remains mathematically cumbersome and restricted to idealized contexts. By directly minimizing the global prediction error on synthetic data, our XAI framework naturally integrates the combined effects of realistic noise, signal correlations, and physical constraints. Furthermore, this model-agnostic approach bypasses the need for complex analytical Jacobians, making it easily adaptable to future models relying on numerical formulations or complex noise distributions.

A striking validation of this approach is the convergence we observed between the data-driven XAI selection and the theoretical weighted CRLB optimization. When the theoretical framework was properly normalized (using A-optimality on relative errors) to account for the disparate units of NEXI parameters, it selected a protocol nearly identical to the one identified by our XGBoost-SHAP-RFE pipeline. This convergence is significant for two reasons. First, it confirms that the XAI effectively learned the underlying information geometry of the biophysical model purely from simulated signal statistics. Second, it suggests that for the NEXI model in this acquisition regime, the presence of non-linear degeneracies does not fundamentally alter the optimal acquisition strategy. This mutual validation strengthens the credibility of the proposed 8-feature protocol, identifying it as a robust global optimum rather than an artifact of a specific selection algorithm.

This theoretical convergence raises a valid question regarding the added value of the XAI pipeline if it ultimately yields the same result as the CRLB. The primary advantage of the XAI approach lies in its model-agnostic flexibility. While calculating the CRLB is highly effective for analytical models like standard NEXI under Gaussian noise, it becomes mathematically cumbersome or even entirely intractable for advanced numerical models lacking analytical Jacobians, or when dealing with complex, non-Gaussian noise distributions, like Rician noise. In such scenarios, the XAI framework serves as a scalable alternative that relies purely on empirical prediction error rather than theoretical derivatives. In the context of this study, the XAI method acts as an independent benchmark, proving that machine-learning-based pruning can achieve analytical gold standards while remaining generalizable to future, more complex modeling frameworks.

The SHAP-RFE framework's analysis revealed that the most informative features are located at the extremes of the ($b$, $\Delta$) acquisition space. This is consistent with the need to sample a wide range of physical regimes to accurately constrain the biophysical model and prevent parameter degeneracy. This distinction proved crucial: standard heuristic or unweighted theoretical approaches often failed to capture the importance of including the longest diffusion time with the strongest b-value. This specific data point, which proved crucial for the accurate estimation of these parameters, was correctly identified as essential by our methodology.

It is worth noting that alternative optimization strategies, such as those proposed by \citet{coelhoAssessmentPrecisionAccuracy2024} or \citet{jespersenOptimizedDiffusionProtocol2025}, could also be considered. These methods typically involve leaving a small, fixed number of acquisition points free to converge iteratively towards optimal positions in continuous parameter space (e.g., b-value). However, for NEXI, where we optimize across ($b$,$\Delta$), this approach risks generating a large number of optimal points, each with a unique diffusion time. Since each unique $\Delta$ value often requires a separate sequence launch on the scanner, this diversity introduces a significant additional operational burden for the researcher or clinician running the protocol. Similarly, arbitrary b-values also limit the usability of the data for mainstream analyses such as DTI/DKI fits, for which a consistent choice of b-values improves cross-site harmonization substantially. By contrast, our SHAP-RFE framework selects a minimal subset from an existing, logistically constrained set of acquisition parameters, and, trained on synthetic NEXI signals spanning the full physiological range and realistic SNR regimes, yielded a reduced protocol that converged with the theoretical optimum derived from weighted CRLB while consistently outperforming expert heuristic strategies. This data-driven approach was chosen because it allows for a comprehensive and uniform sampling of the NEXI parameter space, which is critical for robust and unbiased training of the XGBoost regressor. The selected features span the full b-value range and preserve all $\Delta$ values, supporting estimation of time-dependent kurtosis. Notably, these combinations would be difficult to derive from intuition or analytic simplifications alone.

Our results confirm and extend the utility of SHAP-based RFE pipelines previously applied in other biomedical contexts \citep{zhuIntegratedApproachFeature2025, rufinoPerformanceExplainabilityFeature2024, wangCouplingInterpretableFeature2025}. In particular, similar approaches using XGBoost-SHAP-RFE for feature pruning in physiological state classification \citep{huangMultilayerStackingMethod2024}, symptom triage \citep{rufinoPerformanceExplainabilityFeature2024}, and environmental sensor optimization \citep{wangCouplingInterpretableFeature2025} have demonstrated strong generalization under high-dimensional, noisy conditions or characteristics shared with diffusion MRI signal modeling.

In our context, the XGBoost-SHAP-RFE optimization led to an inflection point at 7 features, beyond which accuracy gains plateaued (Fig. \ref{fig:figure2}). However, deliberately retaining 8 features ensured coverage of both short and long diffusion times, and allowed compatibility with secondary metrics such as time-dependent diffusion kurtosis \citep{henriquesMicroscopicAnisotropyMisestimation2019}. Critically, the final protocol not only reproduced parameter estimates, but also preserved cortical contrast patterns observed in full acquisitions, including elevated $t_{ex}$ and $f$ in sensorimotor cortex \citep{uhlQuantifyingHumanGray2024, chanVivoHumanNeurite2025}.

Finally, it is important to explicitly acknowledge the conceptual limitations of our chosen optimization strategy. By performing feature elimination on a predefined 15-feature master protocol, our framework identifies the optimal subset conditional on these available candidates, rather than solving a global joint optimization problem. In practice, MRI sequence design imposes coupled constraints: pushing for higher b-values or longer diffusion times requires a longer TE, which inherently reduces the base SNR of the entire acquisition due to $T_2$ relaxation. By operating within a fixed master protocol where a single global TE is already established (e.g., TE = 66 ms), our approach treats the SNR and timing of each shell as fixed and independently selectable. It does not account for the potential SNR gains that could be achieved if eliminating the most demanding shells allowed for a sequence-wide TE reduction. 
Here we chose to operate under fixed TE to keep the problem centered on the diffusion-weighting parameter space. Indeed, SNR is also affected by other acquisition parameters such as spatial resolution, coils, acceleration factors, etc. The results of an optimization across a continuous parameter search (as explored in our high-density grid simulation in Supplementary Figure \ref{fig:supp8}) suggests that the optimal points are still on the edges of the $(b, \Delta)$ space accessible given a certain gradient hardware. Thus, restricting the search to a discrete, standard master protocol does not result in a very local optimum, while it ensures that the retained shells remain practical for implementing the protocol at the scanner and perfectly compatible with widespread multi-shell analytical techniques like DTI and DKI.

\subsection{Reproducibility and generalization of the reduced protocol}

Beyond estimation accuracy, test-retest reproducibility is essential for longitudinal studies, particularly when subtle microstructural changes are of interest. We found that the XAI-reduced protocol maintained intra-subject consistency across repeated scans, with only marginal increases in limits of agreement for $t_{ex}$ and $D_i$ compared to the full protocol (Fig. \ref{fig:figure8}). This aligns with prior findings that exchange-related metrics display moderate SNR sensitivity \citep{jelescuNeuriteExchangeImaging2022, uhlQuantifyingHumanGray2024}, but confirms that performance remains within ranges reported in prior studies on clinical and research scanners \citep{uhlQuantifyingHumanGray2024, uhlHumanGrayMatter2025, chanVivoHumanNeurite2025}.

Importantly, the reduced protocol also demonstrated its generalizability to higher-resolution settings in an exploratory 1.6 mm isotropic acquisition performed on one subject. Acquiring data at this finer spatial scale presents a significant challenge due to the inherent reduction in SNR per voxel. To counteract this, we implemented a compensation strategy by substantially increasing the total number of diffusion directions to 1,000. In addition, the increased number of slices resulted in extending the repetition time (TR = 4100 ms) and thus further the scan time. While this necessarily resulted in a four-fold increase in scan duration to 69 minutes, this was a direct consequence of the physical trade-off required to maintain image quality: the voxel volume decreases by a factor of 2, necessitating a four-fold increase in the number of diffusion directions to compensate for the lost signal-to-noise ratio. This remains the optimal NEXI protocol for this high spatial resolution. Indeed, the equivalent full protocol would have lasted 138 minutes. The resulting parameter maps exhibited consistent topography and sharper gray-white matter boundaries compared to the 2 mm scans, suggesting that the XAI-optimized scheme remains robust enough to be extended to finer spatial scales and holds promise for future applications such as laminar-specific analyses at yet higher resolutions. These findings align with recent work advocating for resolution-adaptable diffusion protocols in cortex-focused imaging \citep{hertanuQuantifyingFeaturesHuman2023, palomboSANDICompartmentbasedModel2020}.

Together, these results validate the use of explainable machine learning not merely as a black-box tool, but as a transparent and interpretable strategy for optimizing biophysically-informed acquisition design. The flexibility of our framework allows retraining under alternative models (e.g., SMEX \citep{olesenDiffusionTimeDependence2022}, GEM \citep{uhlGEMUnifyingModel2024}), scanner configurations, or populations of interest, thereby supporting future clinical translation and cross-site harmonization efforts.

\subsection{Spatial specificity and reproducibility across the cortical ribbon}

The spatial distributions of NEXI parameters obtained from the XAI-optimized protocol closely match those observed with the full acquisition, demonstrating that a data-driven subprotocol can preserve biologically relevant contrasts across the cortex. Elevated $t_{ex}$ and $f$ in sensorimotor cortex, previously linked to dense myelination and high neurite packing \citep{uhlQuantifyingHumanGray2024, uhlHumanGrayMatter2025, chanVivoHumanNeurite2025}, were clearly recapitulated. Interestingly, our data partially diverged from earlier studies by showing moderate $t_{ex}$ in visual cortex. This discrepancy may reflect increased sensitivity to fast exchange enabled by the short diffusion times of Connectome 2.0 ($\Delta = 15-45$ ms, $\delta = 5$ ms) \citep{huangConnectome20Developing2021, chanVivoHumanNeurite2025}. We also observed unexpectedly higher $f$ values in temporal regions, despite prior reports associating the temporal lobe with lower cellular and axonal density \citep{ribeiroHumanCerebralCortex2013}, and previous NEXI maps consistently showing the opposite trend in this region \citep{uhlQuantifyingHumanGray2024, uhlHumanGrayMatter2025, chanVivoHumanNeurite2025}. This may be explained by the lower SNR in these regions, which qualitatively affects the robustness of the parameter estimation.

In terms of estimation accuracy, the XAI-optimized protocol demonstrated a substantial improvement over the alternative schemes. RMSD calculations showed that the XAI protocol yielded a 2.4-fold lower RMSD in $t_{ex}$ compared to the Heuristic Mid-Range protocol, and avoided the severe instability in $D_i$ observed with the Heuristic Corner protocol (RMSD 0.11 vs 0.79 µm$^2$/ms). The XAI protocol consistently showed low RMSD and bias across all parameters, comparable to the theoretical CRLB optimum (Figure \ref{fig:figure6}). Importantly, test-retest analyses (Fig.\ \ref{fig:figure8}) confirm that this compact protocol retains measurement reliability for longitudinal studies and sensitivity to between-subject differences. While $t_{ex}$ and $D_i$ showed marginally broader limits of agreement, their reproducibility remained within ranges reported in previous Connectome and Prisma studies \citep{uhlQuantifyingHumanGray2024, uhlHumanGrayMatter2025}. The stability of $f$ and $D_e$ was particularly robust, supporting the feasibility of deploying this protocol in clinical cohorts where subject motion or compliance imposes strict scan-time constraints.

\subsection{Limits of model fidelity and parameter identifiability}

Our analysis also underscores persistent challenges in NEXI parameter estimation. Specifically, $D_i$ often hits its upper model bound ($3.5$ µm$^2$/ms), a behavior also noted in prior work on white matter models like NODDI \citep{jelescuDegeneracyModelParameter2016, howardEstimatingAxialDiffusivity2022} and in recent cortical applications of NEXI \citep{uhlQuantifyingHumanGray2024}, suggesting that this lack of sensitivity to high intra-neurite diffusivity is a general feature of multi-compartment diffusion models. Importantly, while our synthetic simulations (Fig. \ref{fig:figure3}) indicated a tendency for the estimator to underestimate $D_i$ at intermediate values, the systematic overestimation observed in vivo is likely driven by biological confounds not captured in the simulations, such as partial volume effects with cerebrospinal fluid, structural disorder or the presence of unmodeled soma compartments. Similar concerns apply to $t_{ex}$ estimation, where we observe systematic overestimation at intermediate values (Fig.\ \ref{fig:figure3}), likely arising from the interplay between limited short-$\Delta$ data and degeneracy in the fitting landscape. Despite these identifiability challenges at relatively short diffusion times ($\Delta \leq 45$ ms), retaining the exchange component is essential to preserve the model's primary biophysical value: probing membrane permeability. Our results indicate that the XAI-optimized protocol successfully maximizes the available information to constrain $t_{ex}$ within these hardware limitations.

Such biases raise important considerations for model selection and parameter interpretation. As shown in both simulation and experimental studies, models that neglect Rician noise can overestimate $t_{ex}$ \citep{uhlQuantifyingHumanGray2024, chanVivoHumanNeurite2025}, while neglecting unmodeled factors such as cell somas or structural disorder can introduce biases in the estimation of $D_i$, $f$ and even $t_{ex}$ \citep{leeVivoObservationBiophysical2020, novikovExchangeStructurallydisorderedCompartments2023}. These observations echo similar issues reported in white matter modeling \citep{jelescuDegeneracyModelParameter2016, dhitalIntraaxonalDiffusivityBrain2019} and point toward the need for more comprehensive signal models capable of handling gray matter complexity. The pronounced peaks at the parameter boundaries observed in the voxel-wise distributions (Fig. \ref{fig:figure4}), specifically at the $3.5$ \textmu m$^2$/ms limit for $D_i$ and the bounds for $t_{ex}$, reflect these inherent modeling degeneracies. The inclusion of the highest b-value shell ($12.5$ ms/\textmu m$^2$) in both the XAI and CRLB optimal protocols perfectly exemplifies this tension: while it provides critical leverage for constraining $t_{ex}$, it simultaneously renders $D_i$ more vulnerable to Rician noise floors, pushing it toward the upper bound. The XAI optimization navigates this necessary trade-off by minimizing the impact of these boundary effects on the overall normalized error, ensuring that key parameters like $f$ and $D_e$ remain robustly estimated.

Similarly, the finite duration of gradient pulses is known to inflate $D_i$ estimates, an issue addressed by numerical formulations like SMEX \citep{olesenDiffusionTimeDependence2022, uhlHumanGrayMatter2025}. While our preliminary evaluations with SMEX on this dataset highly conserved spatial trends and mitigated the $D_i$ inflation, we retained the analytical NEXI formulation for this methodological study. The ultra-high gradient performance of the Connectome 2.0 system allows for very short gradient durations ($\delta = 5$ ms), substantially mitigating the violation of the Narrow Pulse Approximation compared to clinical systems. Furthermore, the analytical Jacobian of the standard NEXI model provides the necessary numerical stability for rigorous CRLB benchmarking, avoiding the computational complexity of numerical solvers. Consequently, given the hardware capabilities and our focus on robust protocol comparison, standard NEXI provides a stable and sufficient approximation.

\subsection{Denoising tradeoffs and noise-model mismatch}

Another key finding relates to the influence of signal-domain processing. Despite theoretical optimality, complex-valued MP-PCA denoising with Gaussian noise assumptions resulted in poorer model fits at high b-values than magnitude-domain denoising followed by Rician bias correction (Fig.\ \ref{fig:supp4}), a result also seen in \citet{uhlQuantifyingHumanGray2024}. This may reflect residual violations of noise model assumptions, especially in regions with low SNR or spatially correlated noise, arising from different reconstruction steps such as coil combination and parallel imaging \citep{henriquesEfficientPCADenoising2023}.

Future studies may benefit from incorporating more advanced denoising strategies, such as NORDIC \citep{moellerNOiseReductionDIstribution2021} or denoising models with spatially varying noise priors \citep{henriquesEfficientPCADenoising2023}, tailored specifically for cortical data. The possibility of using real-valued reconstructions to suppress Rician floor effects while preserving SNR remains a promising avenue for enhancing estimation robustness.

\subsection{Clinical translation and cross-platform generalizability}

While the proposed 8-feature protocol performs robustly on Siemens Connectome 2.0, its direct deployment on most clinical MRI systems is limited by hardware constraints. For example, Siemens Prisma-class scanners (maximum 80 mT/m) require longer $\delta$ and $\Delta$ to reach high b-values, reducing sensitivity to short-time exchange processes and violating the NPA that underlies the NEXI model. However, the emergence of clinical-grade scanners with ultra-high gradient performance, such as the GE Magnus (300 mT/m, 750 T/m/s) or the Siemens Cima.X (200 mT/m, 200 T/m/s), presents new opportunities. These systems are capable of running protocols similar to the one we've optimized, demonstrating that our data-driven approach is broadly applicable and not limited to a single research scanner. Prior work has shown that on standard clinical scanners, the SMEX model, which integrates full gradient waveforms, yields more accurate estimates \citep{uhlHumanGrayMatter2025, olesenDiffusionTimeDependence2022}.

Nevertheless, the XGBoost-SHAP-RFE optimization framework introduced here can be readily adapted to generate optimized subprotocols under these constraints. By re-training the XGBoost regressors on synthetic signals generated with Prisma-specific pulse timings and noise characteristics, it is possible to derive context-appropriate protocols that preserve the strengths of our approach while ensuring model validity. One of our previous works already begun applying SHAP-based optimization for clinical NEXI acquisition \citep{uhlOptimizingNEXIAcquisition2023}, and our new findings further validate the utility of this pipeline for rapid, interpretable protocol design.

\subsection{Toward multi-model and population-specific optimization}

Finally, we note that the SHAP-RFE approach is model-agnostic and supports extensions beyond NEXI. As the field moves toward multi-compartment models that incorporate soma (e.g., SANDIX) \citep{olesenDiffusionTimeDependence2022, uhlGEMUnifyingModel2024}, structural disorder \citep{novikovExchangeStructurallydisorderedCompartments2023}, or relaxation \citep{leeVivoObservationBiophysical2020}, protocol optimization strategies must evolve accordingly. Our method provides a principled route for empirically selecting features that maximize sensitivity across competing models or prioritize specific biological hypotheses. For instance, this approach could inform the development of free-waveform protocols aiming to enhance sensitivity to both cortical exchange and restriction, complementing existing frameworks such as that of \citet{chakwiziraDiffusionMRIFree2023}.

Furthermore, future work may incorporate priors from pathology-specific cohorts or link optimization targets to downstream biomarkers (e.g., myelin water fraction, cortical thinning, clinical scores). The interpretability and modularity of SHAP facilitate such integration, offering an avenue for personalized or disease-aware protocol design that bridges methodological rigor with translational relevance. Beyond this, our optimization framework could be extended to other acquisition dimensions. For instance, future work could aim to optimize the number of diffusion directions per (b, $\Delta$) combination. This would require a different pipeline than the one used for feature elimination, as simply removing a direction adds noise to the corresponding data point, which could bias the importance ranking and lead to unstable results.

\bigskip

Taken together, our findings demonstrate that XGBoost-SHAP-RFE-based protocol optimization is a powerful and adaptable tool for accelerating diffusion MRI acquisitions while preserving biological specificity. By empirically identifying informative $(b,\Delta)$ pairs under realistic noise and model uncertainty, we provide a scalable framework for optimizing complex biophysical imaging protocols. This framework is generalizable to other scanner platforms, populations, and research questions, and our work represents a key first step towards its broader application. It contributes a key advancement toward the widespread deployment of exchange-sensitive microstructural imaging in research and clinical applications.

% ========= Conclusion =========

\section{Conclusion}

We present a data-driven protocol optimization framework for Neurite Exchange Imaging (NEXI), leveraging explainable machine learning (XGBoost-SHAP-RFE) to identify a compact 8-feature acquisition scheme that maintains estimation accuracy, spatial specificity, and repeatability. Applied on the Connectome 2.0 platform, this reduced protocol cuts scan time nearly by half compared to the full 15-feature reference, while matching the performance of a rigorous CRLB optimization and significantly outperforming expert heuristic strategies.

Our results demonstrate a striking convergence between data-driven feature selection and theoretical A-optimality, confirming that XAI can effectively identify the information geometry of biophysical models directly from simulated signals. Crucially, the optimized protocol avoids the pitfalls of intuitive designs: it eliminates the estimation bias observed with dense mid-range sampling and mitigates the noise instability inherent to extreme corner sampling. This balance is particularly vital for robustly estimating challenging parameters like $t_{ex}$ and $f$ in the presence of realistic noise.

While not directly portable to clinical systems due to gradient constraints, the proposed optimization method is model-agnostic and can be retrained for clinical MRI scanner platforms using realistic simulations. This flexible framework represents a critical step toward clinically viable, exchange-sensitive dMRI protocols for probing cortical microstructure in vivo.

\section*{Data and code availability}

The code used in this study is available at \url{https://github.com/Mic-map/graymatter_swissknife}. All code for XGBoost-SHAP-RFE optimisation is publicly available at \url{https://github.com/QuentinUhl/XAI-dMRI-protocol-optimization}. 
The data that support the findings of this study are available from the corresponding author upon reasonable request and will be publicly deposited in an OpenNeuro repository (\url{https://openneuro.org}) upon publication of this article.
Access is unrestricted under the CC-BY-4.0 licence.

\section*{Author contributions}

Conceptualization: IJ, QU; \\
Data acquisition: QU, JG, BB, YJ, SF, AB; \\
Data curation: QU;  \\
Formal analysis: QU, KC, YM; \\
Funding acquisition: IJ, BB, SH; \\ 
Investigation: QU;  \\
Methodology: QU,  IJ, BB;  \\
Supervision: IJ, BB, SH, HL; \\
Visualization: QU, TP, QW; \\
Writing - original draft: QU;  \\
Writing - review \& editing: all authors; \\

\section*{Funding}

QU, TP and IJ were supported by the Swiss National Science Foundation (SNSF)  
Eccellenza grant PCEFP2\_194260.  

Work performed at the Athinoula A.\ Martinos Center, Massachusetts General Hospital, 
was supported by the National Institutes of Health under the following grants:  
R01 EB028797, R01 EB032378, R01 EB034757, R21 AG082377, R21 EB036105,  
P41 EB030006, U01 EB026996, and UG3 EB034875.

KC is supported by NWO/ZonMw under the Rubicon award number 04520232330012.

QU gratefully acknowledges the Swiss National Science Foundation (SNSF) for a Mobility Grant that enabled his research visit to the Athinoula A. Martinos Center for Biomedical Imaging, Massachusetts General Hospital.

\section*{Declaration of competing interests}

The authors declare that they have no known competing financial interests or personal relationships that could have appeared to influence the work reported in this paper.

\bibliographystyle{apalike}
\bibliography{better_bibtex_paper3_r1_v0}

\clearpage
\appendix
\vspace*{3em}
\begin{center}
    {\huge \textbf{Supplementary Materials}} \\
    \vspace{1.5em}
    {\Large \textbf{Reduced NEXI protocol for the quantification of human gray matter microstructure on the Connectome 2.0 scanner}} \\
    \vspace{1.5em}
    
    % --- Authors list ---
    {\large
    Quentin Uhl\textsuperscript{1}~\orcidlink{0000-0002-9160-4605},
    Tommaso Pavan\textsuperscript{1}~\orcidlink{0000-0002-3436-6882},
    Julianna Gerold\textsuperscript{2}~\orcidlink{0009-0000-5460-6591},
    Kwok-Shing Chan\textsuperscript{2}~\orcidlink{0000-0001-8427-169X}, \\ \vspace{0.3em}
    Yohan Jun\textsuperscript{2}~\orcidlink{0000-0003-4787-4760},
    Shohei Fujita\textsuperscript{2}~\orcidlink{0000-0003-4276-6226},
    Aneri Bhatt\textsuperscript{2}~\orcidlink{0009-0003-3502-9399},
    Yixin Ma\textsuperscript{2}~\orcidlink{0000-0002-2080-9013}, \\ \vspace{0.3em}
    Qiaochu Wang\textsuperscript{1}~\orcidlink{0009-0007-0234-6864},
    Hong-Hsi Lee\textsuperscript{2}~\orcidlink{0000-0002-3663-6559},
    Susie Y. Huang\textsuperscript{2}~\orcidlink{0000-0003-2950-7254}, \\ \vspace{0.3em}
    Berkin Bilgic\textsuperscript{2,*}~\orcidlink{0000-0002-9080-7865},
    Ileana Jelescu\textsuperscript{1,*}~\orcidlink{0000-0002-3664-0195}
    } \\
    \vspace{1.5em}
    
    % --- Affiliations ---
    {\small
    \textsuperscript{1} Department of Radiology, Lausanne University Hospital (CHUV) and University of Lausanne, Lausanne, Switzerland \\ \vspace{0.2em}
    \textsuperscript{2} Department of Radiology, Massachusetts General Hospital, Athinoula A. Martinos Center for Biomedical Imaging, Boston, MA, USA \\ \vspace{0.2em}
    \textsuperscript{*} Joint last authorship
    }
\end{center}
\vspace{3em}

% Supplementary document preamble
%\counterwithin*{section}{document}
\renewcommand{\thesection}{S\arabic{section}}
\renewcommand{\thesubsection}{S\arabic{section}.\arabic{subsection}}
\renewcommand{\thefigure}{S\arabic{figure}}
\renewcommand{\thetable}{S\arabic{table}}
\setcounter{section}{0}
\setcounter{subsection}{0}
\setcounter{figure}{0}
\setcounter{table}{0}
% ---------- Additional Figures from Main Paper Supplementary ----------
% ---------- Figure S1 ----------
\begin{figure}[H]
\centering
\includegraphics[width=0.80\textwidth]{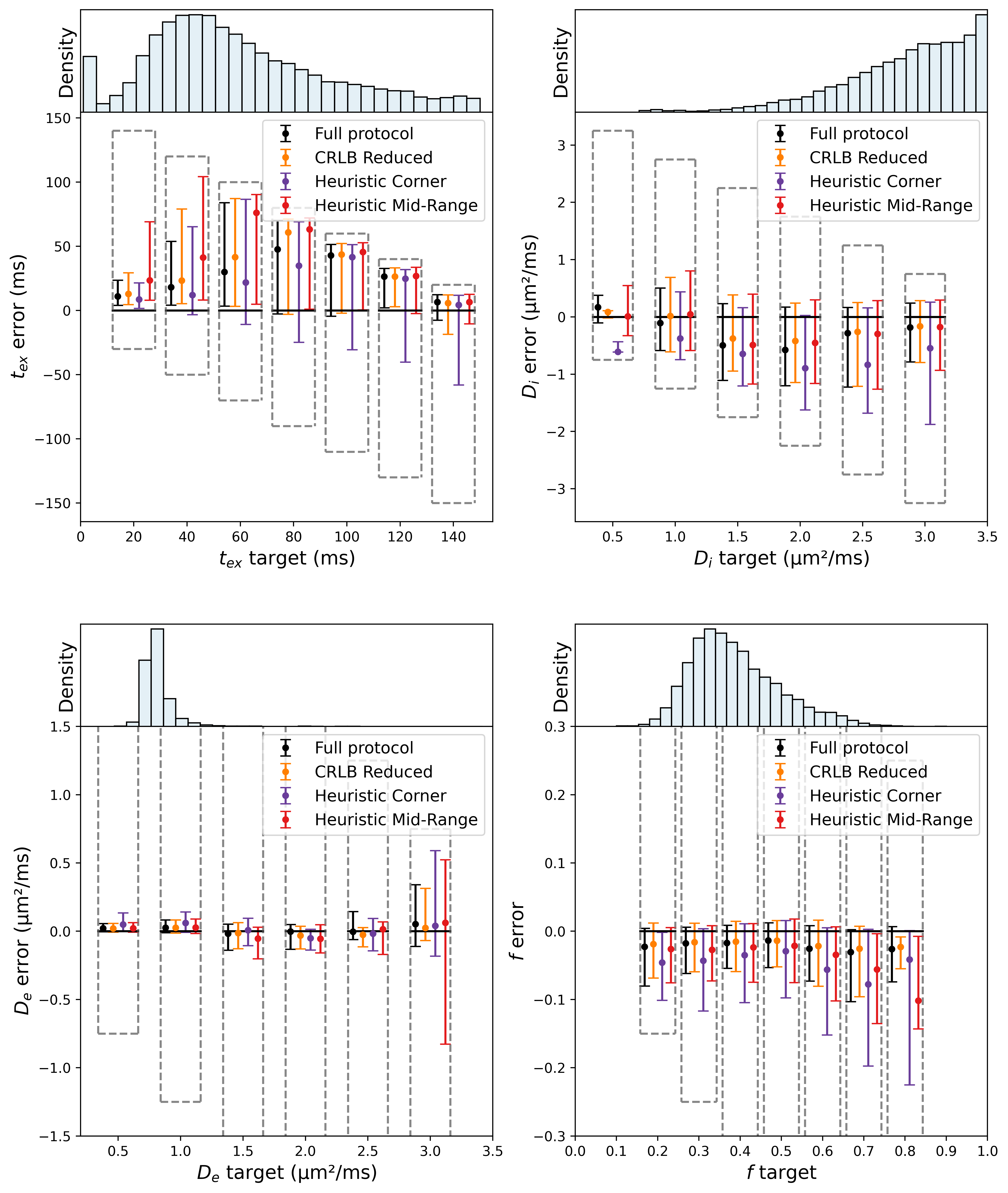}
\caption{
\textbf{Estimation errors from synthetic simulations: Full vs. CRLB and Heuristic subprotocols.}
Boxplots show absolute errors in parameter estimation ($t_{ex}$, $f$, $D_i$, $D_e$) for synthetic NEXI signals corrupted with Gaussian noise (SNR=32$\pm$26). As in Figure 3, y-axis limits are scaled per parameter to improve visibility. Dashed boxes represent model fitting bounds.
The Heuristic Mid-Range protocol (red) shows the largest errors for $t_{ex}$, confirming its poor sensitivity to exchange due to limited $\Delta$ sampling.
The Heuristic Corner protocol (purple) exhibits increased variance for $D_i$, consistent with its in vivo instability.
Both the CRLB (orange) and XAI (see Fig. \ref{fig:figure3}) protocols maintain errors closer to the Full protocol baseline.
}
\label{fig:supp1}
\end{figure}
% ---------- Figure S2 ----------
\begin{figure}[H]
\centering
\includegraphics[width=0.80\textwidth]{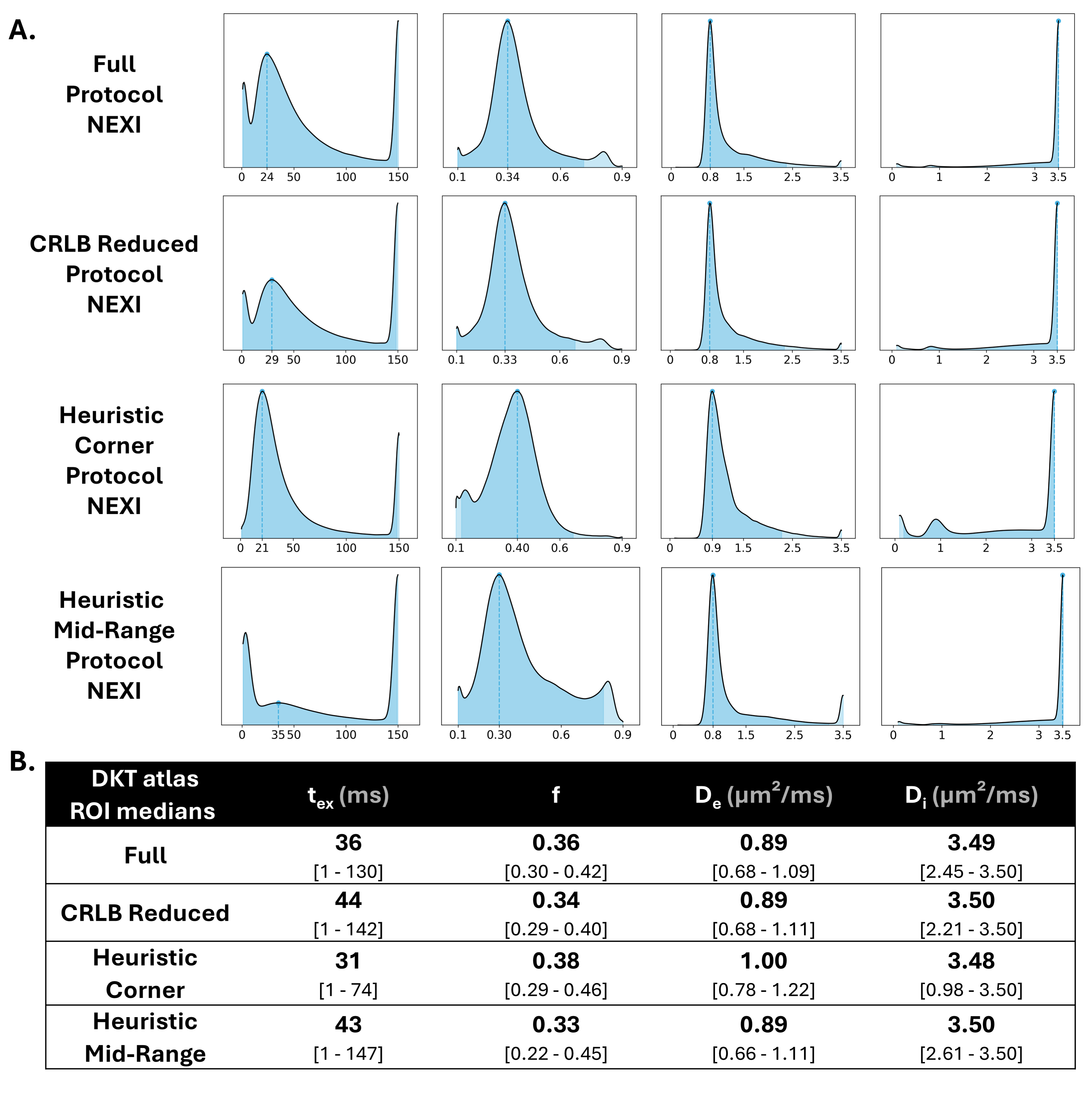}
\caption{
\textbf{Distributions and summary of NEXI parameter estimates across DKT cortical regions for full, CRLB, and heuristic protocols.}
(A) Kernel density estimates of voxel-level values for each NEXI parameter ($t_{ex}$, $f$, $D_e$, $D_i$), aggregated across all DKT ROIs and all subjects. The Full protocol (top row) serves as the reference, displaying characteristic unimodal distributions for $f$ and $D_e$, and a right-skewed distribution for $t_{ex}$.
The CRLB Reduced protocol (second row) closely mirrors the reference distributions, confirming its theoretical optimality.
In contrast, the Heuristic Corner protocol (third row) exhibits a distorted distribution for $D_i$ with a spurious secondary peak at low diffusivity values, reflecting the estimation instability caused by noise sensitivity at high b-values.
The Heuristic Mid-Range protocol (bottom row) shows a sharpened and shifted distribution for $t_{ex}$ toward the boundaries, indicative of the systematic bias caused by insufficient diffusion time sampling.
The vertical dashed lines indicate the identified mode of each distribution.
(B) Summary statistics of ROI-wise medians. While these aggregate metrics reveal specific biases, such as the inflated $D_e$ for the Heuristic Corner protocol ($1.00$ vs $0.89$ µm$^2$/ms), they largely mask the severe estimation instability observed in the distributions. Crucially, the median $D_i$ for the Heuristic Corner protocol appears deceptively close to the reference, failing to reflect the substantial variance and distributional distortion visible in (A).
}
\label{fig:supp2}
\end{figure}
% ---------- Figure S3 ----------
\begin{figure}[H]
\centering
\includegraphics[width=\textwidth]{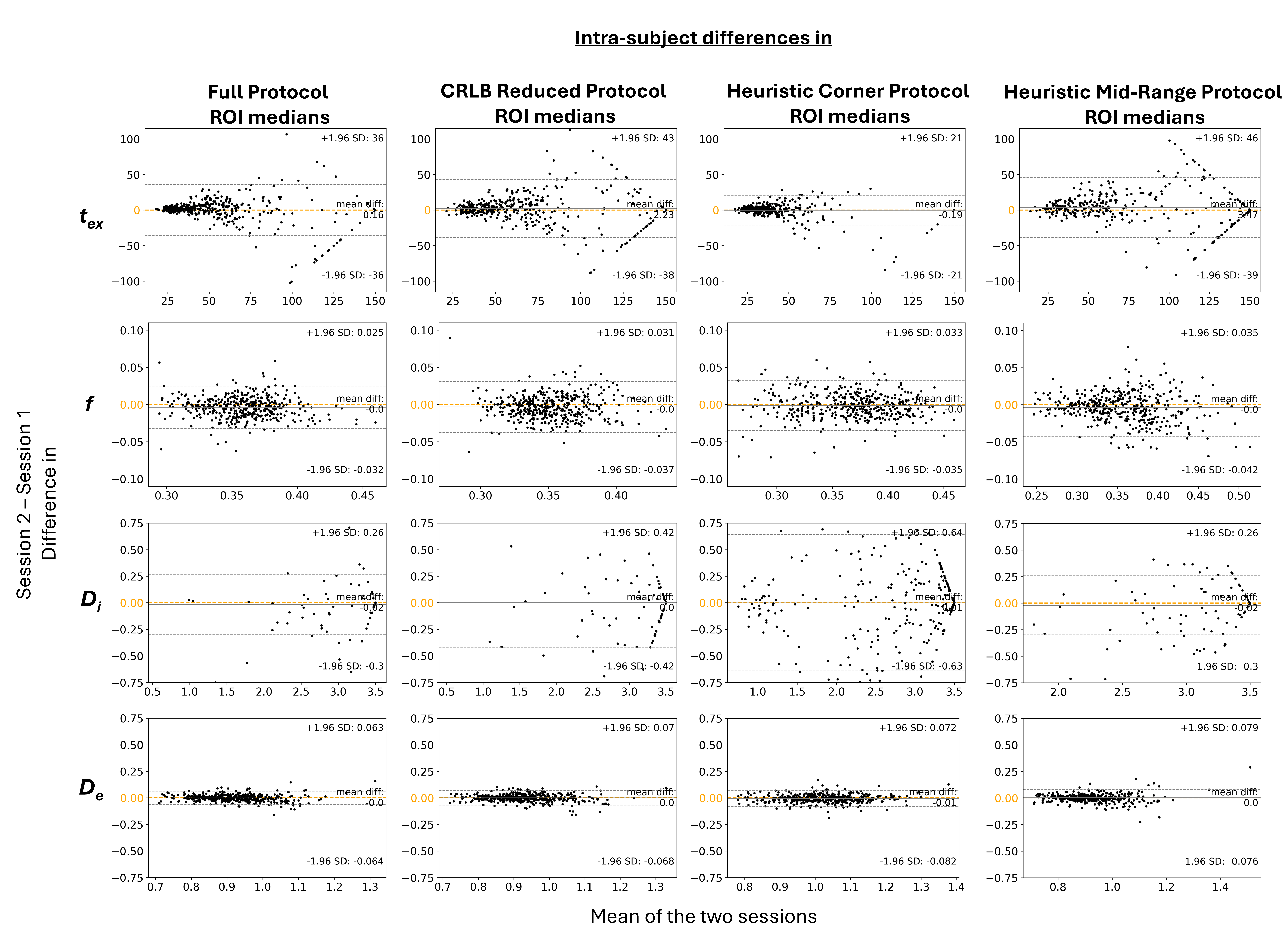}
\caption{
\textbf{Test–retest reproducibility of full, weighted-CRLB, and heuristic subprotocols.}
Bland–Altman plots show intra-subject differences in DKT ROI medians between repeated sessions for $t_{ex}$, $f$, $D_i$, and $D_e$.
Overall, all reduced protocols maintain a reproducibility generally comparable to the Full protocol (left) for stable parameters like $f$ and $D_e$. However, critical differences emerge for exchange-sensitive metrics.
Remarkably, the Heuristic Corner protocol (third column) demonstrates the best reproducibility for $t_{ex}$, with tighter limits of agreement ($\pm 21$ ms) than even the Full protocol ($\pm 36$ ms). This highlights the benefit of maximizing the geometric lever arm for constraining exchange parameters. However, this strategy fails for $D_i$, where noise sensitivity leads to marked instability (Limits: $\pm 0.64$ vs $\pm 0.26$ µm$^2$/ms for Full).
The Heuristic Mid-Range protocol (right) and CRLB Reduced protocol (second column) show balanced reproducibility profiles comparable to the Full protocol, without the instability observed in the Corner approach.
}
\label{fig:supp3}
\end{figure}

% ---------- End of Additional Figures from Main Paper Supplementary ----------
% ---------- Cross scanner Supplementary ----------
\clearpage
\subsection*{Supplementary Comparison Across Scanners and Pipelines}
To contextualize our current results on the CONNECTOM 2.0 scanner, we compared group-level NEXI parameter estimates to those obtained in our previous studies using both CONNECTOM 1.0 and PRISMA scanners. The CONNECTOM 1.0 dataset comprised data from four healthy adults (Age: 30.5±3.8 years) acquired on a 3T scanner with 300 mT/m gradients. The acquisition protocol used b-values from 1.0 to 7.5 ms/µm$^2$, at four $\Delta$ from 20 to 49 ms, with a fixed gradient duration of $\delta$=9 ms. The protocol totaled 700 volumes and 45 minutes of scan time \cite{uhlQuantifyingHumanGray2024}. The PRISMA dataset included data from 11 healthy adults (Age: 26.9±1.3 years) scanned on a 3T system with 80 mT/m gradients. This protocol, requiring a total of 27 minutes, sampled b-values from 1.0 to 5.0 ms/µm$^2$ at five $\Delta$ from 28.3 to 65.0 ms, with a fixed gradient duration of $\delta$=16.5 ms and 20 directions per shell, totaling 325 volumes \cite{uhlHumanGrayMatter2025}. The two supplementary tables below summarize values across cortical regions using either the mode of means (Table S1) or the mean of medians (Table S2), allowing fair comparison across acquisition hardware, denoising strategies, and modeling variants (e.g., standard NEXI vs. SMEX). Notably, $t_{ex}$ is consistently longer and $D_i$ reaches upper bounds more frequently in the CONNECTOM 2.0 data, likely reflecting its higher gradient strength and sensitivity to fast intra-neurite dynamics. At the same time, inter-scanner discrepancies in $f$ and $D_e$ remain modest. These cross-platform results highlight both the potential and challenges of interpreting NEXI parameters across hardware configurations and processing choices.
% ---------- Table S1----------
\begin{table}[h!]
\centering
\setlength{\fboxsep}{10pt}
\fbox{%
\begin{tabular}{@{}lcccc@{}}
\toprule
 & $t_{\text{ex}}$ (ms) & $D_i$ ($\mu m^2/ms$) & $D_e$ ($\mu m^2/ms$) & $f$ \\
\midrule
CONNECTOM 2.0 (Complex)       & \textbf{55.7} & \textbf{3.26} & \textbf{1.12} & \textbf{0.37} \\
                              & [6.1 -- 116.6] & [2.35 -- 3.50] & [0.84 -- 1.41] & [0.29 -- 0.44] \\
\addlinespace
PRISMA NEXI (Magnitude)       & \textbf{41.1} & \textbf{2.61} & \textbf{1.59} & \textbf{0.51} \\
                              & [15.5 -- 66.7] & [2.03 -- 3.20] & [1.18 -- 2.00] & [0.41 -- 0.61] \\
\addlinespace
PRISMA SMEX (Wide Pulses)     & \textbf{36.8} & \textbf{2.45} & \textbf{1.23} & \textbf{0.42} \\
                              & [17.6 -- 58.0] & [1.99 -- 2.90] & [0.88 -- 1.61] & [0.34 -- 0.50] \\
\bottomrule
\end{tabular}
}
\vspace{0.5em}
\caption{ Summary of NEXI parameter estimates using the \textit{mode of the means} across DKT cortical regions.\\
This comparison includes data from CONNECTOM 2.0 (with complex-valued denoising), and PRISMA (magnitude-based denoising), processed using either standard NEXI or SMEX (a wide-pulse variant). Results show similar $f$ and $D_e$ values across systems, while $t_{ex}$ is notably longer and $D_i$ consistently higher in CONNECTOM 2.0, frequently reaching the upper model bound. These trends suggest enhanced sensitivity to fast exchange processes at shorter diffusion times, but also point toward possible model limitations.}
\label{tab:suppt1}
\end{table}
\vspace{2em}
% ---------- End Table S1 ----------
% ---------- Table S2 ----------
\begin{table}[h!]
\centering
\setlength{\fboxsep}{10pt}
\fbox{%
\begin{tabular}{@{}lcccc@{}}
\toprule
 & $t_{\text{ex}}$ (ms) & $D_i$ ($\mu m^2/ms$) & $D_e$ ($\mu m^2/ms$) & $f$ \\
\midrule
CONNECTOM 1.0 (NEXI$_{\text{RM}}$) & \textbf{42.3} & \textbf{3.35} & \textbf{0.92} & \textbf{0.38} \\
                                 & [40.0 -- 44.7] & [3.32 -- 3.38] & [0.91 -- 0.93] & [0.379 -- 0.389] \\
\addlinespace
CONNECTOM 2.0 (Complex)          & \textbf{57.9} & \textbf{3.40} & \textbf{0.92} & \textbf{0.36} \\
                                 & [23.2 -- 150.0] & [2.26 -- 3.50] & [0.75 -- 1.12] & [0.30 -- 0.41] \\
\bottomrule
\end{tabular}
}
\vspace{0.5em}
\caption{ Comparison of NEXI parameter estimates using the \textit{mean of DKT medians} with 95\% confidence intervals (CI).\\
Estimates are derived from two acquisition protocols: CONNECTOM 1.0 using magnitude denoising with the Rician mean model (NEXI$_{\text{RM}}$), and CONNECTOM 2.0 using complex-valued denoising. While $f$ and $D_e$ remain comparable between systems, $t_{ex}$ is substantially higher in CONNECTOM 2.0. Additionally, $D_i$ values in C2 reach the model’s upper fitting bounds in many ROIs, potentially reflecting sensitivity to fast intra-neurite diffusion or the lack of explicit soma modeling.}
\label{tab:suppt2}
\end{table}
% ---------- End Table S2 ----------
% ---------- End of Cross scanner Supplementary ----------
% ---------- Denoising Impact Supplementary ----------
\clearpage
\subsection*{Impact of denoising strategy on signal modeling and parameter estimation}
\begin{figure}[H]
\centering
\includegraphics[width=\textwidth]{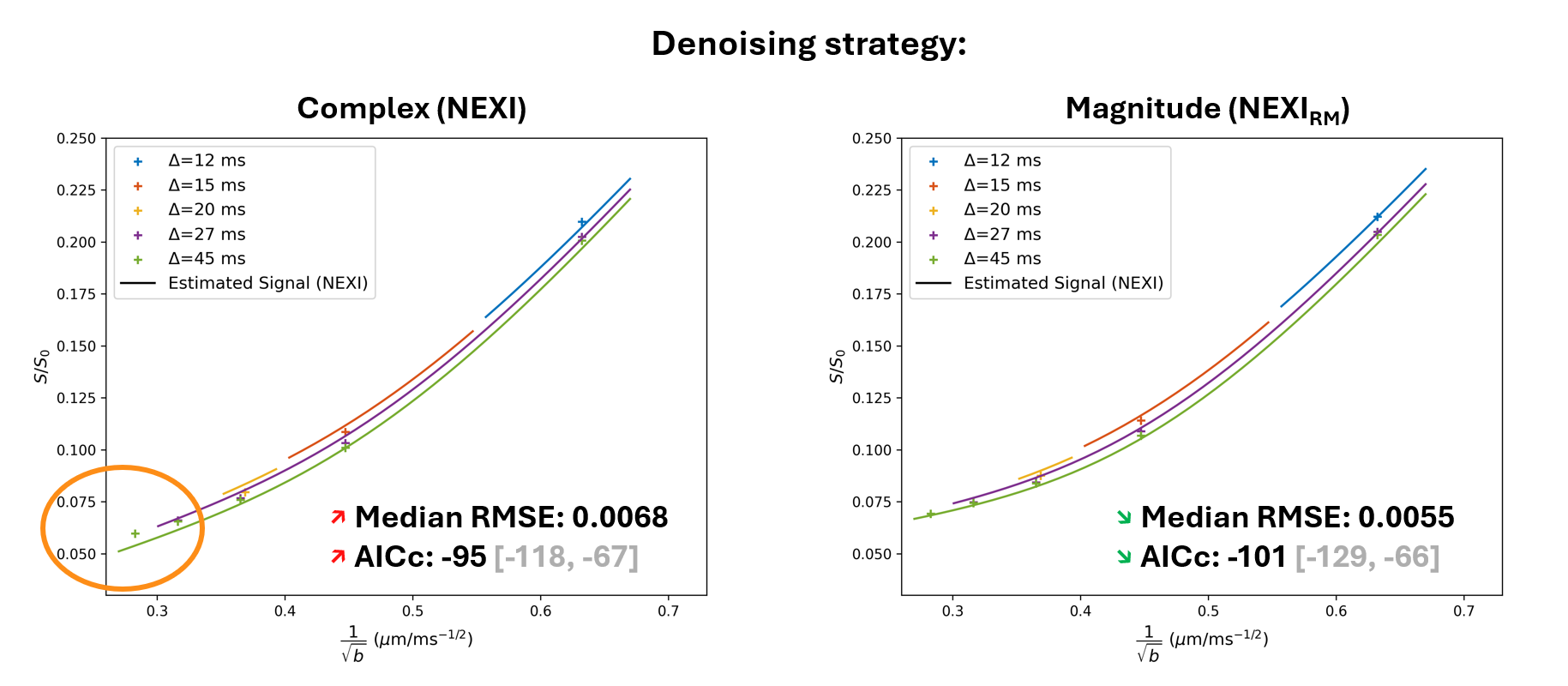}
\caption{
\textbf{Comparison of NEXI model fitting using complex vs. magnitude-denoised data.}
Mean cortical diffusion signals are plotted as a function of $1/\sqrt{b}$ for each diffusion time $\Delta$ (color-coded), along with the corresponding NEXI model fit (black lines). Note that lower $1/\sqrt{b}$ values correspond to the highest diffusion weightings. 
Left: Complex-valued MP-PCA denoising followed by standard NEXI fitting. 
Right: Magnitude-based denoising followed by Rician-mean corrected NEXI fitting (NEXI$_\text{RM}$). 
Although complex denoising is theoretically less biased, it yields a higher residual median RMSE (0.0068) and worse AICc ($-95$) compared to the magnitude-based pipeline (RMSE = 0.0055, AICc = $-101$). 
Notably, a visible mismatch appears for high b-values (i.e., low $1/\sqrt{b}$) and long diffusion times ($\Delta = 45$ ms, green) in the complex case, suggesting a deviation between signal and model fit that is not present in the magnitude-based pipeline. 
This difference may reflect a residual model inadequacy at high diffusion weighting or subtle SNR-driven effects. 
These results underscore the need to empirically evaluate model–data agreement even under theoretically optimal denoising conditions.
}
\label{fig:supp4}
\end{figure}

Although complex-valued denoising is often favored in diffusion MRI for its minimal bias under Gaussian noise assumptions, its superiority for model fitting is not guaranteed in practice. In this section, we compare model–data agreement and parameter estimates between two processing pipelines: (i) complex-valued MP-PCA denoising with standard NEXI fitting, and (ii) magnitude denoising with Rician mean correction (NEXI$_\text{RM}$). To ensure statistical robustness, all models were fitted at the voxel level; the signals and curves shown hereafter represent the average of these individual voxel-wise fits across the cortical ribbon. Figure \ref{fig:supp4} evaluates model fit quality against mean cortical signals. This result emphasize the importance of empirically validating denoising choices, especially under conditions of high diffusion weighting or limited SNR, rather than relying solely on theoretical bias arguments.
% ---------- Figure S4 ----------

% ---------- End of Denoising Impact Supplementary ----------
% ---------- High resolution Supplementary ----------
\clearpage
\subsection*{Extension to High-Resolution Acquisition}
\begin{figure}[H]
\centering
\includegraphics[width=\textwidth]{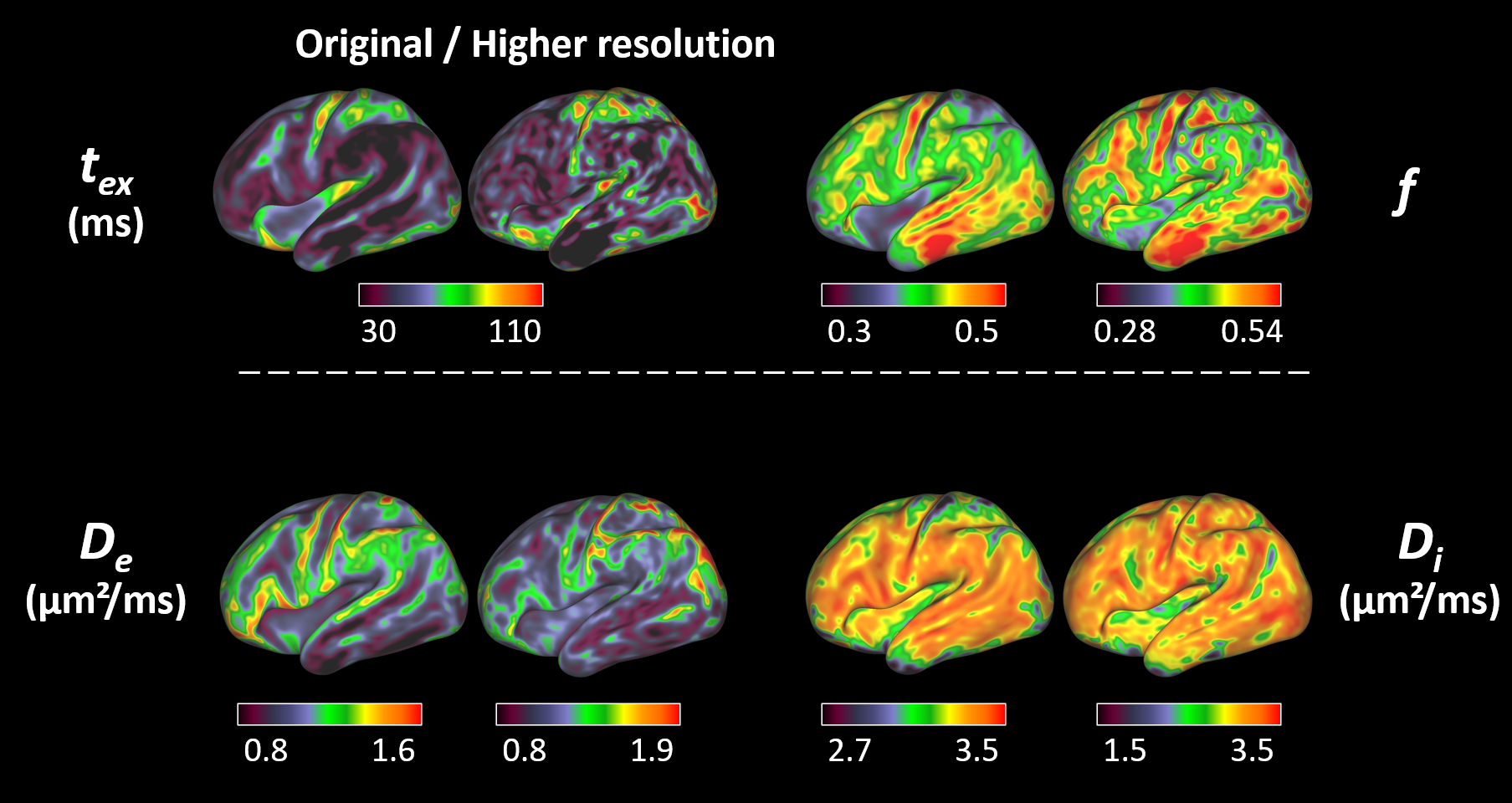}
\caption{
\textbf{Surface-based comparison of NEXI parameters at standard (2.0 mm) and high (1.6 mm) resolution.}
Spatial distributions are largely preserved across resolutions, confirming the robustness of parameter estimation under increased spatial detail. Notably, subtle differences emerge in fine anatomical structures, particularly in $f$ and $t_{ex}$, suggesting improved delineation of regional heterogeneity at higher resolution. Slight increases in parameter range (as seen in colorbars) may reflect reduced partial volume effects and enhanced anatomical specificity.
This confirms that the reduced 8-feature protocol remains robust at higher spatial resolution and may support laminar or fine-grained microstructural analyses in longer acquisitions.
}
\label{fig:supp5}
\end{figure}

\begin{figure}[H]
\centering
\includegraphics[width=\textwidth]{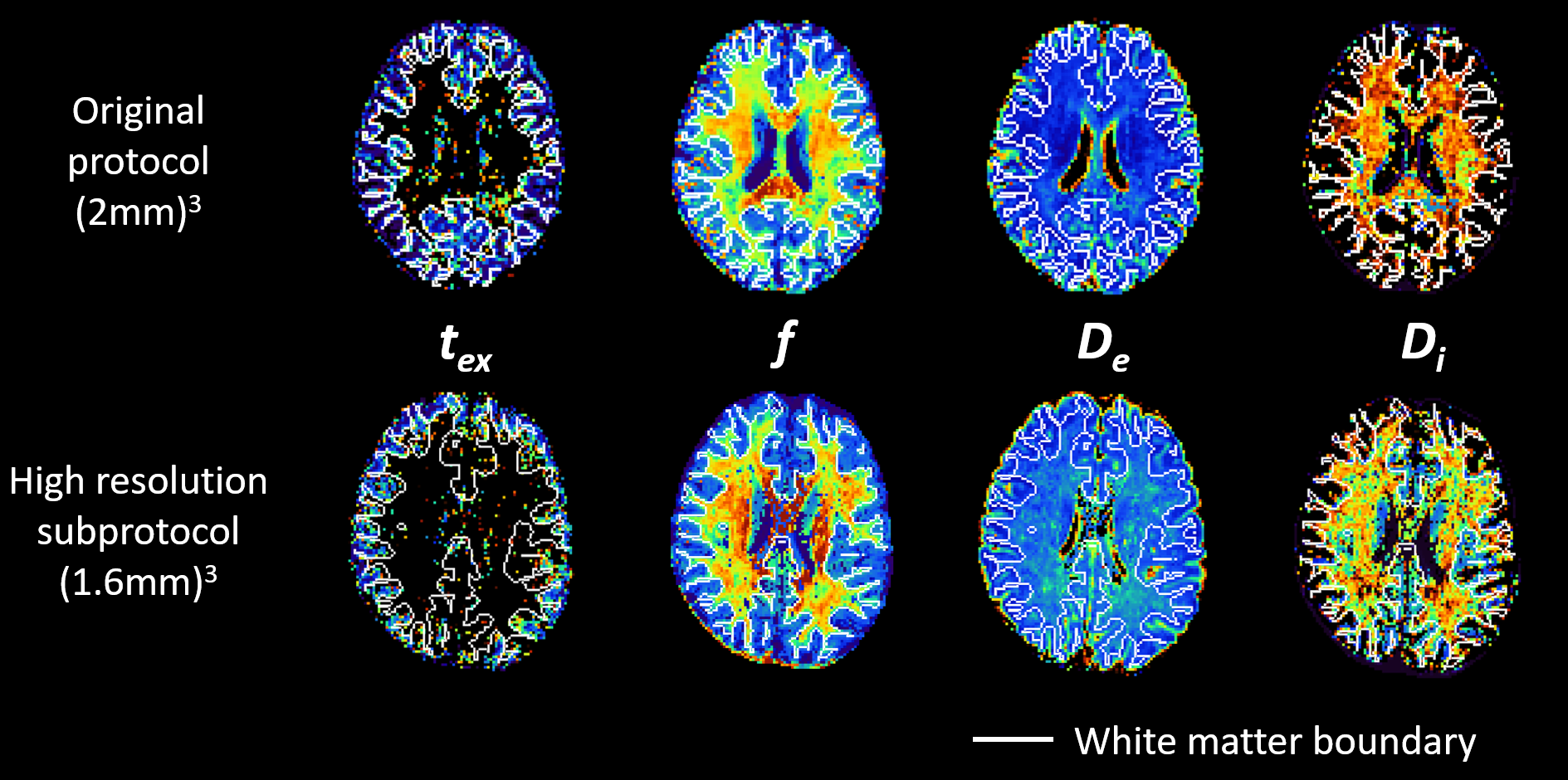}
\caption{
\textbf{Axial slice comparison of NEXI parameter maps at standard (2.0 mm) and high (1.6 mm) isotropic resolution.}
Maps of $t_{ex}$, $f$, $D_e$, and $D_i$ are displayed across the same mid-cortical axial slice in a single subject, comparing the original full protocol (top row) to the high-resolution version of the reduced 8-feature protocol (bottom row). Despite the smaller voxel size, parameter estimates remain qualitatively consistent, confirming the protocol’s robustness across spatial resolutions. 
High-resolution maps reveal sharper gray–white matter boundaries (highlighted in white), improved delineation of cortical ribbon structure, and increased local contrast, especially for $f$ and $D_i$. These enhancements suggest that the reduced protocol supports fine-grained anatomical mapping, and may be suitable for studies targeting laminar profiles or subtle cortical microstructural changes.
}
\label{fig:supp6}
\end{figure}

Beyond designing a clinically viable protocol, we also assessed its performance in a high-resolution setting to address the challenge of Partial Volume Effects (PVE), which commonly confound microstructural measurements in cortical gray matter. One subject was scanned using the 8-feature optimized protocol at 1.6 mm isotropic resolution over a 60-minute session. The resulting NEXI parameter maps remained qualitatively consistent with those obtained at 2 mm resolution, as shown both on the cortical surface (Fig. \ref{fig:supp5}) and in axial slices (Fig. \ref{fig:supp6}). High-resolution acquisitions provided sharper delineation of the cortical ribbon and improved contrast at the gray-white matter boundary, particularly for $f$ and $D_i$, with no indication of increased instability in model fitting. These results demonstrate that our reduced protocol can be leveraged for more accurate and reliable cortical mapping when extended scan durations are feasible.

% ---------- Figure S5 ----------
% ---------- Figure S6 ----------
% ---------- End of High resolution Supplementary ----------
% ---------- Comparison with weighted-FIM Supplementary ----------
\clearpage
\subsection*{Convergence of Data-Driven and Theoretical Optimizations}

In the main text, we demonstrated that the XAI-optimized protocol converges with the theoretical Weighted A-optimality (CRLB) criterion. To provide a comprehensive theoretical validation and ensure our results are not dependent on a single loss function, we also evaluated the Weighted D-optimality criterion based on the Fisher Information Matrix (FIM).

While A-optimality (used in the main text via the CRLB) minimizes the average relative variance, D-optimality seeks to minimize the volume of the confidence ellipsoid by maximizing the determinant of the information matrix. To ensure consistency with the weighted CRLB approach and account for the disparate units of NEXI parameters, we implemented a weighted FIM optimization defined as:
\[
\max \det(\mathbf{J}^T \mathbf{J})
\]
where $\mathbf{J}$ represents the weighted Jacobian, normalized by both the noise standard deviation associated with each acquisition and the ground-truth parameter values. This formulation aligns with the relative error minimization objective of the w-CRLB and XAI approaches.

\begin{figure}[H]
    \centering
    \includegraphics[width=\textwidth]{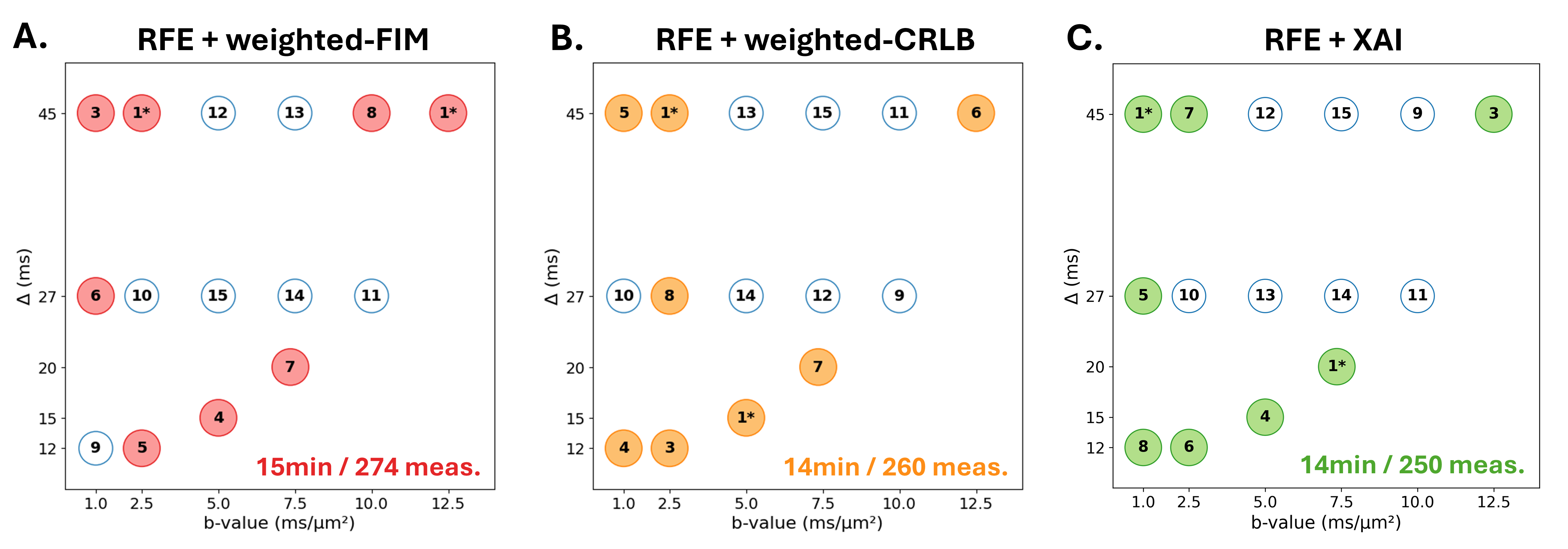}
    \caption{\textbf{Convergence of D-optimality (FIM) with A-optimality (CRLB) and XAI.} 
    Comparison of the final feature sets selected by: \textbf{(A)} Weighted FIM RFE (D-optimality), \textbf{(B)} Weighted CRLB RFE (A-optimality, used in the main text), and \textbf{(C)} XAI-based RFE. 
    Colored circles indicate the ($b, \Delta$) combinations retained in the final 8-feature protocols, while empty circles represent features eliminated during the RFE process. The numbers inside the circles correspond to the shell indices of the full 15-feature protocol. 
    All three methods independently converge towards a "corner-based" strategy, selecting the highest and lowest available $b$-values and diffusion times to maximize sensitivity. The FIM-optimized protocol confirms the trend observed with CRLB and XAI, validating that the chosen acquisition scheme is robust regardless of the specific theoretical optimality criterion applied.}
    \label{fig:supp7}
\end{figure}

As illustrated in Figure \ref{fig:supp7}, the protocol resulting from this Weighted D-optimality optimization (Panel A) is strikingly similar to both the w-CRLB (Panel B) and XAI (Panel C) protocols presented in the main text. Specifically, the FIM-optimized protocol prioritizes the exact same "corners" of the sampling space ($b_{min}/b_{max}$ and $\Delta_{min}/\Delta_{max}$) and shares 7 out of 8 features with the XAI protocol. This triple convergence across distinct optimization strategies, minimizing prediction error (XAI), minimizing parameter variance (CRLB), and maximizing information volume (FIM), strongly supports the global optimality of the selected 8-feature protocol.

% ---------- Figure S7 ----------

% ---------- End of Comparison with weighted-FIM Supplementary ----------

% ---------- High-Density Grid Simulation Supplementary ----------
\clearpage
\subsection*{Extended Optimization on a High-Density Continuous Grid}

To demonstrate the robustness of our data-driven XAI pipeline and its capacity to scale beyond the initial 15-feature master protocol, we extended the optimization to a highly dense acquisition grid. A grid of 100 theoretical $(b, \Delta)$ combinations was generated. To ensure the physical viability of this search space, an exclusion mask was applied to strictly limit the maximum achievable $b$-value for any given $\Delta$ according to the hardware constraints of the Connectome 2.0 scanner ($G_{max} = 500$ mT/m, $\delta = 5$ ms). This filtering left 80 physically realizable $(b, \Delta)$ combinations. The number of diffusion directions for each synthetic shell was scaled following the heuristic rule $N_{dir} \propto \sqrt{b}$ (constrained between 20 and 64).

\begin{figure}[H]
    \centering
    \includegraphics[width=0.90\textwidth]{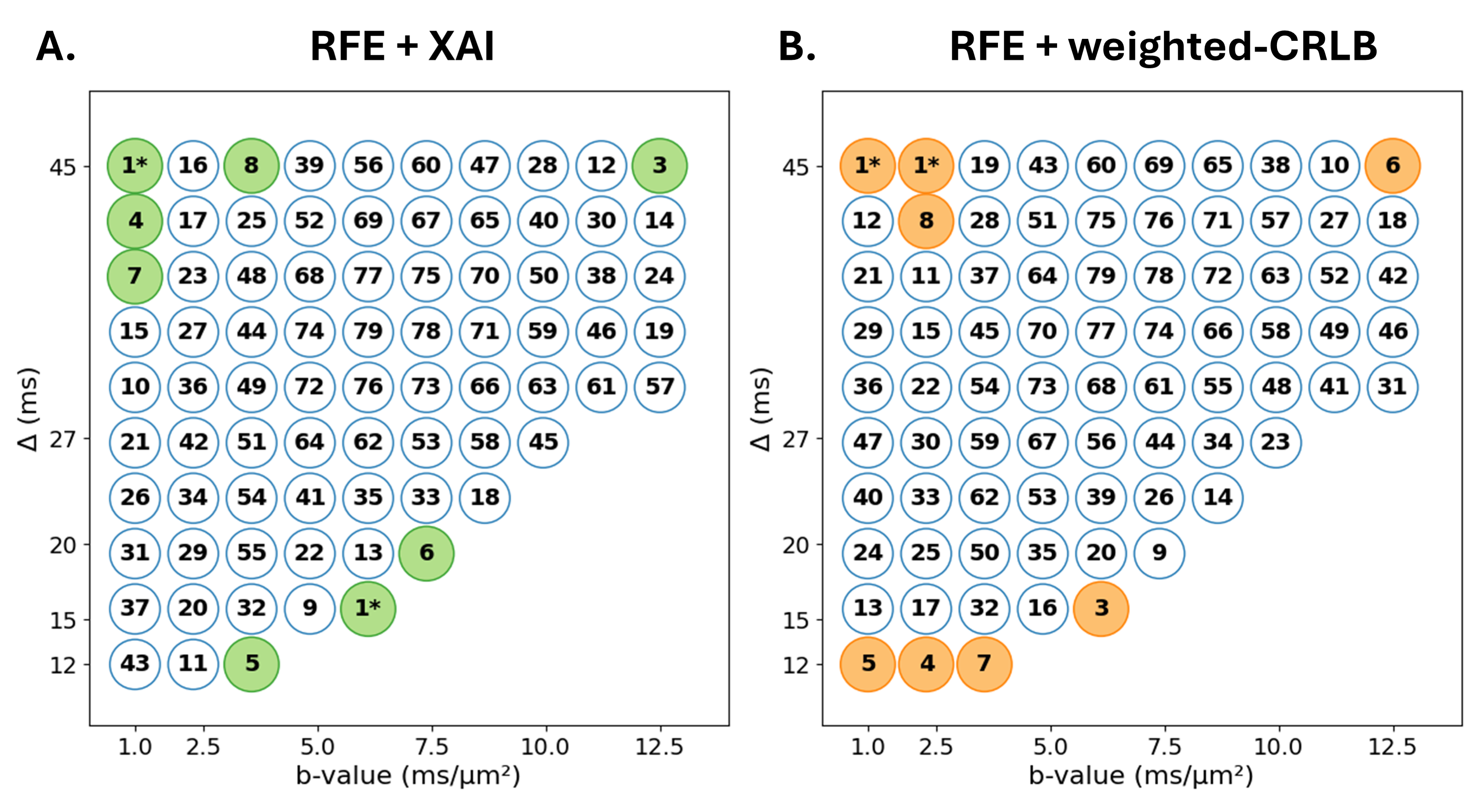}
    \caption{\textbf{Final optimized sub-protocols on a high-density, physically constrained grid.} 
    XAI (left) and weighted-CRLB (right) optimizations were performed on an initial pool of 80 physically realizable $(b, \Delta)$ combinations. The numbers inside the circles indicate the elimination rank, with lower numbers representing the most important, last-remaining features. Both methods converge toward a similar macroscopic acquisition strategy, primarily retaining features (highlighted in green and orange) located at the extreme edges of the parameter space rather than along the maximum gradient efficiency diagonal. This confirms that the underlying information geometry relies heavily on boundary conditions, which is consistent with the subset selected from the 15-feature master protocol in the main text.}
    \label{fig:supp8}
\end{figure}

Both the XGBoost-SHAP-RFE pipeline and the weighted-CRLB optimization were executed on this dense 80-feature space. Figure \ref{fig:supp8} illustrates the final retained features. Remarkably, both the data-driven and theoretical optimizations exhibit a highly similar behavior: they predominantly select features at the extreme boundaries of the physically achievable parameter space, rather than clustering along the maximum gradient diagonal. Notably, the XAI approach maintains a slightly greater temporal diversity across $\Delta$ compared to the CRLB.

While both methods converge toward a common "corner-heavy" protocol geometry, performing recursive elimination on such a dense grid introduces significant multicollinearity. Neighboring $(b, \Delta)$ points provide highly redundant information, making the early elimination steps delicate as many points share a very low individual impact. This observation provides strong post-hoc justification for our decision to restrict the primary in vivo optimization to a well-spaced 15-feature master protocol. By avoiding dense redundancy, the 15-feature approach stabilizes early feature ranking, ensures each point contributes unique information, and implicitly maintains compatibility with standard diffusion models, like DTI or DKI, by relying on conventional b-value shells.

% ---------- Figure S8 ----------
% ---------- End of High-Density Grid Simulation Supplementary ----------

\end{document}